\documentclass[twocolumn,superscriptaddress]{revtex4-1}
\usepackage{amsmath}
\usepackage{graphicx}
\usepackage{dcolumn}
\usepackage{bm}
\usepackage[colorlinks, linkcolor=blue, citecolor=blue, urlcolor=blue,breaklinks=true]{hyperref}
\usepackage{color}
\usepackage[toc,page,header]{appendix}
\makeatletter
\renewcommand\paragraph{\@startsection{paragraph}{4}{\z@}%
   {-3.25ex\@plus -1ex \@minus -.2ex}%
   {1.5ex \@plus .2ex}%
   {\normalfont\normalsize\textit}}
\makeatother

\begin{document}
\title{Converting microwave and telecom photons with a silicon photonic nanomechanical interface}
\author{G.~Arnold}
\email{Authors contributed equally.}
\affiliation{Institute of Science and Technology Austria, 3400 Klosterneuburg, Austria}
\author{M.~Wulf}
\email{Authors contributed equally.}
\affiliation{Institute of Science and Technology Austria, 3400 Klosterneuburg, Austria}
\author{S.~Barzanjeh}
\email{Present address: Institute for Quantum Science and Technology (IQST),
University of Calgary, Canada.}
\affiliation{Institute of Science and Technology Austria, 3400 Klosterneuburg, Austria}
\author{E.~S.~Redchenko}
\affiliation{Institute of Science and Technology Austria, 3400 Klosterneuburg, Austria}
\author{A.~Rueda}
\affiliation{Institute of Science and Technology Austria, 3400 Klosterneuburg, Austria}
\author{W.~J.~Hease}
\affiliation{Institute of Science and Technology Austria, 3400 Klosterneuburg, Austria}
\author{F.~Hassani}
\affiliation{Institute of Science and Technology Austria, 3400 Klosterneuburg, Austria}
\author{J.~M.~Fink}
\email{jfink@ist.ac.at}
\affiliation{Institute of Science and Technology Austria, 3400 Klosterneuburg, Austria}
\date{\today}
\begin{abstract}
Practical quantum networks require low-loss and noise-resilient optical interconnects as well as non-Gaussian resources for entanglement distillation and distributed quantum computation \cite{Kimble2008}. The latter could be provided by superconducting circuits \cite{Blais2004} but - despite growing efforts and rapid progress \cite{Higginbotham2018,Fan2018} - existing solutions to interface the microwave and optical domains lack either scalability or efficiency, and in most cases the conversion noise is not known. In this work we utilize the unique opportunities \cite{Safavi-Naeini2019} of silicon photonics, cavity optomechanics and superconducting circuits to demonstrate a fully integrated, coherent transducer connecting the microwave X and the telecom S bands with a total (internal) 
bidirectional transduction efficiency of 1.2\,\% (135\,\%) at millikelvin temperatures. 
The coupling relies solely on the radiation pressure interaction mediated by the femtometer-scale motion of two silicon nanobeams and includes an
optomechanical gain of about 20\,dB. 
The chip-scale device 
is fabricated from CMOS compatible materials and achieves a $V_\pi$ as low as $16\,\mu\text{V}$ for sub-nanowatt pump powers. 
Such power-efficient, ultra-sensitive and highly integrated hybrid interconnects might find applications ranging from quantum communication~\cite{Stannigel2010} and RF receivers ~\cite{Bagci2014} to magnetic resonance imaging~\cite{Simonsen2019}. 
%
\end{abstract}
\maketitle


Large scale quantum networks will facilitate the next level in quantum information technology, 
such as the internet did for classical communication, enabling e.g. secure communication
and distributed quantum computation~\cite{Kimble2008}.
Some of the most promising platforms to process quantum information locally, such as superconducting circuits~\cite{Blais2004}, spins in solids~\cite{Awschalom2018}, and quantum dots~\cite{Burkard2020},
operate naturally in the gigahertz frequency range, but the long-distance transmission of gigahertz radiation is relatively lossy and not resilient to ambient noise. This limits the length of supercooled microwave waveguides in a realistic scenario to tens of meters. 
In contrast, the transport of quantum information over distances of about 100 kilometers is nowadays routinely achieved by sending optical photons at telecom frequency through optical fibers.

There is a variety of platforms, which in principle have shown to be able to merge the advantages of both worlds 
ranging from mechanical, piezoelectric, electro-optic, magneto-optic, rare-earth and Rydberg atom implementations\,~\cite{Lambert2019,Lauk2019}. 
So far, the optomechanical approach has been proven to be most efficient, reaching a record high photon conversion efficiency of up to 47\,\% with an added noise photon number of only 38~\cite{Higginbotham2018}. But this composite device is based on a Fabry-Perot cavity that has to be hand assembled and utilizes a membrane mode that is restricted to relatively low mechanical frequencies. Using piezo-electricity coherent conversion between microwave and optical frequencies has been shown uni- and bidirectional ~\cite{Vainsencher2016,jiang2019} at room temperature,
and at low temperatures~\cite{Forsch2020,Han2020} with integrated devices, 
so far with either low efficiency or unspecified conversion noise properties. 
The electro-optic platform has shown promising photon conversion efficiencies \cite{Rueda2016}, recently up to 2\,\%~ at 2~K~\cite{Fan2018}, but generally suffers from the need of very high pump powers in the milliwatt regime.

\begin{figure*}[t]
\centering
\includegraphics[width=1.9\columnwidth]{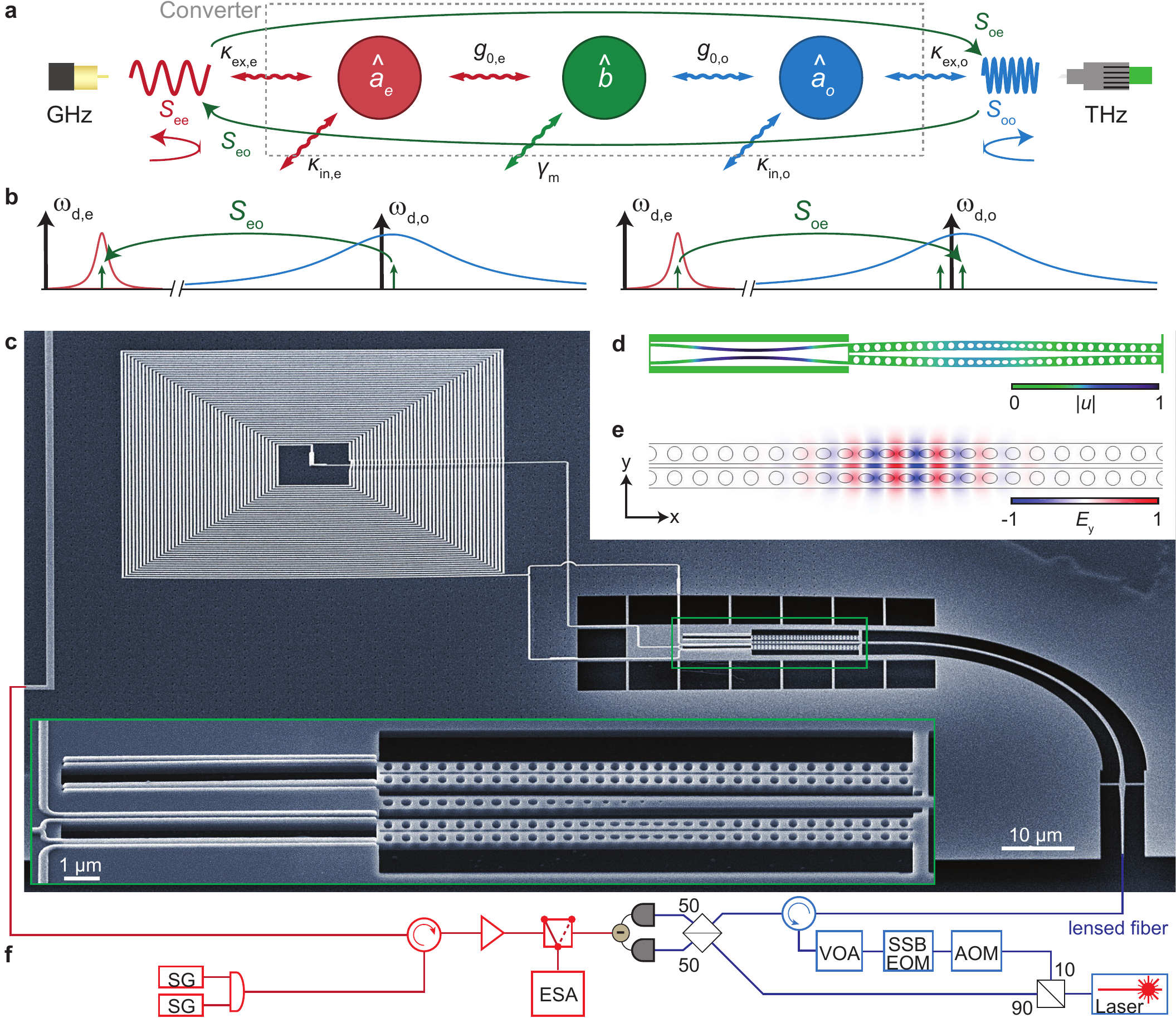}
\caption{
\textbf{Silicon photonic microwave-to-optics converter.} 
\textbf{a}, Diagram showing the microwave $(\hat a_\text{e})$, mechanical $(\hat b)$, and optical $(\hat a_\text{o})$ mode, and the relevant coupling and loss rates of the 
device. Scattering parameters $S_{ij}$ characterizing the transducer performance are indicated.
\textbf{b}, Schematic indicating the coherent signals involved in the conversion process (green). The bidirectional conversion $\zeta$ is only evaluated between the upper sidebands at $\omega_{\text{d},j}+\omega_\text{m}$.    
\textbf{c}, Scanning electron micrograph of the device showing the microwave lumped element resonator with an inductively coupled feed line, the photonic crystal cavity and the optical coupling waveguide fabricated on a fully suspended 220\,nm thick silicon on insulator device layer. The inset shows an enlarged view of the central part (green boxed area) comprising the mechanically compliant vacuum gap capacitors of size $\sim 70\,\text{nm}$ and two optomechancial zipper cavities (top one unused) with a central tapered photonic crystal mirror coupled optical waveguide. 
\textbf{d}, Finite element method (FEM) simulation of the mechanical displacement $|u|$ of the utilized mechanical resonance. 
\textbf{e}, FEM simulation of the electric in-plane-field ${E_\text{y}(x,y)}$ for the relevant optical mode. 
\textbf{e}, Simplified experimental setup. The device is mounted on the mixing chamber plate of a cryogen-free dilution refrigerator at a temperature of $\mathrm{T_\text{fridge}=50\,\text{mK}}$. 
A microwave switch selects between the incoming microwave and optical signal to be analyzed by the ESA. Optical heterodyning is used to detect the low power levels used in the experiment. Acronyms: microwave signal generator (SG), electronic spectrum analyzer (ESA), variable optical attenuator (VOA), single-sideband electro-optic modulator (SSB EOM), acousto-optic modulator (AOM).
}
 \label{Fig_sample_properties}
\end{figure*}

In this work we present a device
that converts coherent signals between 10.5\,GHz and 198\,THz 
at millikelvin temperatures via the radiation pressure interaction, as schematically shown in Fig.~\ref{Fig_sample_properties}a and b. Due to the need of only picowatt range pump powers, both the heat load to the cryostat and local heating of the integrated device is minimized. Fabricated from aluminum on a commercial silicon-on-insulator wafer over an area of approximately $200\,\mu\text{m}\times120\,\mu\text{m}$, see Fig.~\ref{Fig_sample_properties}c,
it is compact, versatile and
fully compatible with silicon photonics 
and superconducting qubits~\cite{Keller2017}. The unique electro-opto-mechanical design is optimized for very strong field confinements and radiation pressure couplings, which enable internal efficiencies exceeding unity for ultra-low pump powers.  
We find that the conversion noise
so far precludes a quantum limited operation and we present a comprehensive theoretical and experimental noise analysis to evaluate the potential for scalable and noise-free conversion in the future. 

The transducer consists of one microwave resonator and one optical cavity, both parametrically coupled via the vacuum coupling rates $g_{\text{0},j}$ with $j=\text{e,o}$ to the same mechanical oscillator as shown in Fig.~\ref{Fig_sample_properties}a. The intrinsic decay rate of the optical (microwave) resonator is $\kappa_{\text{in,o}}\,(\kappa_{\text{in,e}})$, while the optical (microwave) waveguide-resonator coupling is given by $\kappa_{\text{ex,o}}\,(\kappa_{\text{ex,e}})$ resulting in a total damping rate of $\kappa_j=\kappa_{\text{in},j}+\kappa_{\text{ex},j}$ and coupling ratios $\eta_j=\kappa_{\text{ex},j} / \kappa_{j}$. The mechanical oscillator, with intrinsic decoherence rate $\gamma_{\text{m}}$ and frequency $\omega_{\text{m}}$, is shared between the optical cavity and the microwave resonator and acts as a bidirectional coherent pathway to convert the photons between the two different frequencies~\cite{Stannigel2010, Regal2011, Safavi2011, Barzanjeh2012}. In the interaction picture, the Hamiltonian describing the conversion process is (see Supplementary Material)
\begin{eqnarray}
\label{eq_beamsplitHam}
\hat{H}_{\text{int}}=\sum_{j=\text{e,o}}\Big(\hbar G_{j} (\hat{a}_j^\dagger\hat{b}+\hat{a}_j\hat{b}^{\dagger}) +\hat H_{\mathrm{CR},j}\Big),
\end{eqnarray}
where $\hat{a}_j$, $(\hat b)$ with $j=\text{e,o}$ is the annihilator operator of the electromagnetic (mechanical) mode, and $\hat H_{\text{CR},j}= \hbar G_{j} (\hat{a}_j\hat{b}\,e^{2i\omega_\text{m}}+\text{h.c.})$ describes the counter-rotating terms which are responsible for the coherent amplification of the signal. $G_{j}=\sqrt{n_{\text{d},j}}g_{\text{0},j}$ is the parametrically-enhanced electro- or optomechanical coupling rate where $n_{\text{d},j}$ is the intracavity photon number due to the corresponding microwave and optical pump tones. In the resolved-sideband regime $4\omega_\text{m}>\kappa_j$ we neglect $\hat H_{\text{CR},j}$ under the rotating-wave approximation 
and the Hamiltonian (\ref{eq_beamsplitHam}) represents a beam-splitter like interaction in which the mechanical resonator mediates noiseless photon conversion between microwave and optical modes.
%
%
Note that near-unity photon conversion $\zeta_{\text{RS}}=4 \eta_{\text{e}}\eta_{\text{o}} \mathcal{C}_\text{e}\mathcal{C}_\mathrm{o} / (1 +\mathcal{C}_\text{e}+\mathcal{C}_\text{o})^2$
can be achieved 
in the limit of $\mathcal{C}_\text{e}=\mathcal{C}_\text{o}\gg1 $ with $ \mathcal{C}_j=4G_j^2/(\kappa_j \gamma_\text{m})$ being the electro- or optomechanical cooperativity, as demonstrated between two optical~\cite{Hill2012} and two microwave modes~\cite{Lecocq2016, Fink2019}. 

\begin{figure*}[t]
\centering
\includegraphics[width=2\columnwidth]{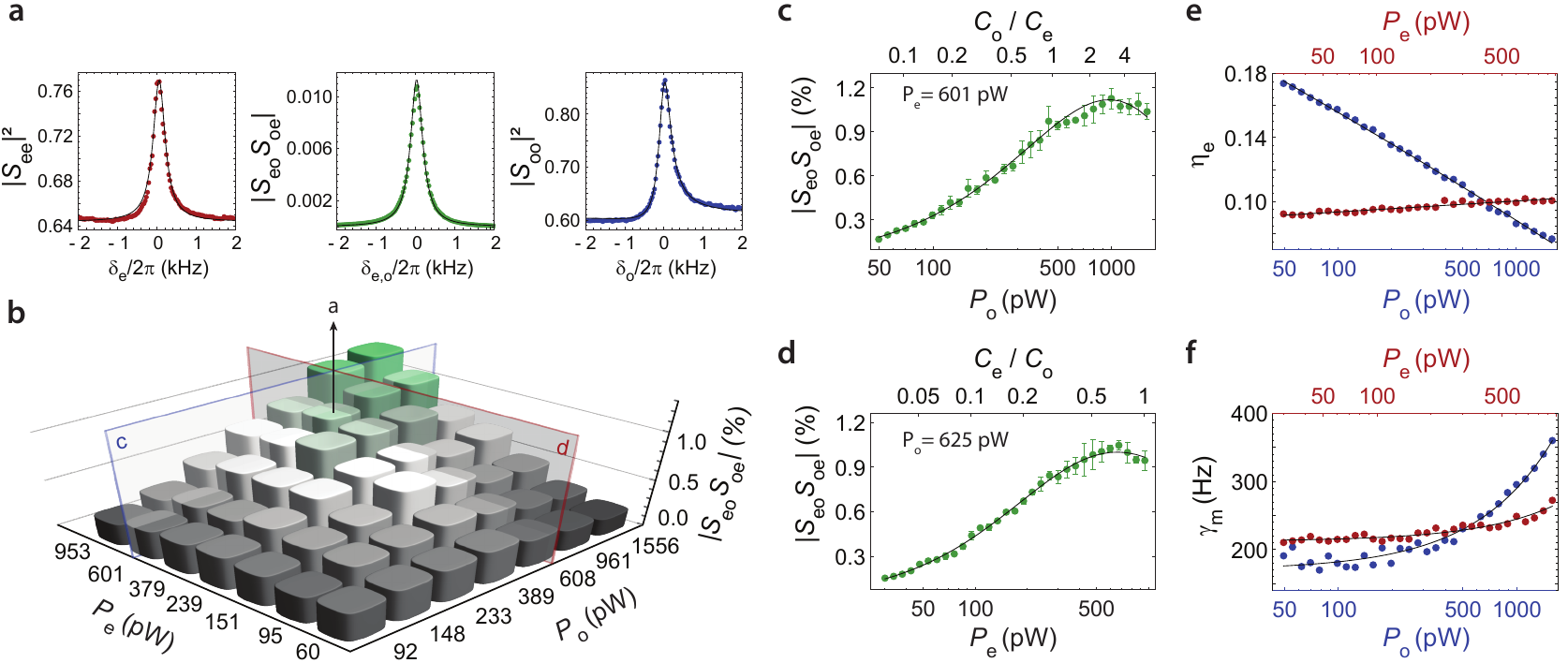}
\caption{\textbf{Coherent scattering parameter measurements.} 
\textbf{a}, The reflection $|S_{jj}|^2$ ($j=\text{e,o}$) and bidirectional transduction $\zeta:=|S_{\text{eo}}S_{\text{oe}}|$ parameters as a function of probe tone detuning $\delta_j=\omega-(\omega_{\mathrm{d},j}+\omega_{\mathrm{m}})$ for fixed pump powers $P_\text{e}=601\,\text{pW}$ and $P_\mathrm{o}=625\,\text{pW}$. The dots represent the experimental data while the solid lines show the theoretical prediction with $\gamma_{\text{m}}$ as the only fit parameter. 
\textbf{b}, Measured photon number conversion efficiency as a function of microwave and optical pump powers. \textbf{c}, Measured conversion efficiency with respect to optical pump power for fixed $P_{\text{e}}=601\,\text{pW}$. \textbf{d}, Measured conversion efficiency with respect to microwave pump power for fixed $P_{\mathrm{o}}=625\,\text{pW}$. In \textbf{c} and \textbf{d}, the error bars are the standard deviation of three independent measurement runs and solid lines are theory with interpolated $\gamma_{\text{m}}$ (from panel \textbf{f}) and no other free parameters. 
\textbf{e}, The coupling efficiency of the microwave resonator to the waveguide $\eta_\text{e}=\kappa_{\text{ex,e}}/\kappa_\text{e}$, extracted from broad band reflection measurements, as a function of optical (blue, $P_{\text{e}}=601\,\text{pW}$) and microwave (red, $P_{\text{o}}=625\,\text{pW}$) pump tones. 
\textbf{f}, Intrinsic mechanical decoherence rate $\gamma_{\text{m}}$ versus optical (blue, $P_{\text{e}}=601\,\text{pW}$) and microwave (red, $P_{\text{o}}=625\,\text{pW}$) pump powers extracted from \textbf{c} and \textbf{d} with Eq.~\ref{eq_transduction} and $\gamma_{\text{m}}$ as only fit parameter. Solid lines are linear fits to the data.}\label{Fig_conversion}
\end{figure*}

We realize conversion by connecting an optomechanical photonic crystal zipper cavity~\cite{Safavi-Naeini2013b} with two aluminum coated and mechanically compliant silicon nanostrings ~\cite{Barzanjeh2019}
as shown in Fig.~\ref{Fig_sample_properties}c. The mechanical coupling between these two components is carefully designed, leading to a hybridization of their in-plane vibrational modes into symmetric and antisymmetric supermodes. In case of the 
antisymmetric mode that is used in this experiment, the strings and the photonic crystal beams vibrate 180 degrees out of phase as shown by the finite-element method (FEM) simulation 
in Fig.~\ref{Fig_sample_properties}d. 
%
The photonic crystal cavity features two resonances at telecom frequencies with similar optomechanical coupling strength. 
The simulated spatial distribution of the electric field component $\mathrm{E_\text{y}(x,y)}$ of the higher frequency mode with lower loss rate used in the experiment is shown in Fig.~\ref{Fig_sample_properties}e.
The lumped element microwave resonator consists of 
an ultra-low stray capacitance planar spiral coil inductor and two mechanically compliant capacitors with a vacuum gap of size of $\sim 70\,\text{nm}$. 
This resonator is inductively coupled to a shorted coplanar waveguide, which is used to send and retrieve microwave signals from the device. 
The sample is fabricated 
using a robust multi-step recipe including electron beam lithography, silicon etching, 
aluminum thin-film deposition and hydrofluoric vapor acid etching, 
as described in detail in Ref.~\cite{Dieterle2016}.

Standard sample characterization reveals a optical resonance frequency of $\omega_{\text{o}}/(2\pi)=198.081\,\text{THz}$ with total loss rate $\kappa_{\text{o}}/(2\pi)=1.6\,\text{GHz}$ and waveguide coupling rate $\kappa_{\text{ex,o}}/(2\pi)=0.18\,\text{GHz}$ leading to a coupling efficiency of $\eta_{\text{o}}=0.11$. When the optical light is turned off, the microwave resonance frequency is $\omega_{\text{e}}/(2\pi)=10.5\,\text{GHz}$ with coupling efficiency $\eta_{\text{e}}=0.4$ and $\kappa_{\text{ex,e}}/(2\pi)=1.15\,\text{MHz}$. The mechanical resonator frequency has a value of $\omega_{\text{m}}/(2\pi)=11.843\,\text{MHz}$ with an intrinsic decoherence rate $\gamma_{\text{m}}/(2\pi)=15\,\text{Hz}$ at a mode temperature of 150\,mK. The achieved single-photon-phonon coupling rates are as high as $g_{\text{0,e}}/(2\pi)=67\,\text{Hz}$ and $g_{\text{0,o}}/(2\pi)=0.66\,\text{MHz}$.


To perform coherent photon conversion, red-detuned microwave and optical tones with powers $P_{\text{e(o)}}$ 
are applied to the microwave and the optical resonator. These drive tones establish the linearized electro- and optomechanical interactions, which results in the conversion of a weak microwave (optical) signal tone to the optical (microwave) domain measured in our setup as shown in Fig.~\ref{Fig_sample_properties}f. We experimentally characterize the transducer efficiency by measuring the normalized reflection $|S_{jj}|^2$ ($j=\text{e,o}$) and the bidirectional transmission $\zeta:=|S_{\text{eo}}S_{\text{oe}}|$ coefficients as a function of probe detuning $\delta$. As shown in Fig.~\ref{Fig_conversion}a, for drive powers $P_{\text{e}}=601\,\text{pW}$ and $P_{\text{o}}=625\,\text{pW}$ with detunings $\Delta_j=\omega_{j}-\omega_{\text{d},j}$ of $\Delta_{\text{e}}=\omega_{\text{m}}$ and $\Delta_{\text{o}}/(2\pi)=126\,\text{MHz}$ leading to intracavity photon numbers of $n_{\text{d,e}} \approx 9\cdot 10^5$ and $n_{\text{d,o}} \approx 0.2$,
the measured total (waveguide to waveguide) photon transduction efficiency is $\approx~1.1\,\%$ corresponding to $96.7\,\%$ internal (resonator to resonator) photon transduction efficiency over the total bandwidth of $\Gamma_{\text{conv}}/(2\pi)\approx0.37\,\text{kHz}$. The probe tone adds 17($10^{-3}$)
 photons to the microwave resonator (optical cavity). Here we use a self-calibrated measurement scheme 
that is independent of the gain and loss of the measurement lines as described in Ref.~\cite{Andrews2014} and we only take into account transduction between the upper two sidebands at $\omega_{\mathrm{d},j}+\omega_{\mathrm{m}}$ as shown in Fig.~\ref{Fig_sample_properties}b.  
Neglecting the lower optical sideband that is generated 
due to the non-sideband resolved situation $\kappa_{\text{o}} / 4\omega_{\text{m}} \approx 30$ reduces the reported bidirectional efficiencies by $\sqrt{2}$ compared to the actually achieved total transduction efficiency between microwave and optical fields.
The observed reflection peaks indicate that both resonators are under-coupled, equivalent to an impedance mismatch for incoming signal light.
All scattering parameters are in excellent agreement with our theoretical model (solid lines, see Supplementary Material) with $\gamma_{\text{m}}$ the only free fit parameter. 

Figure \ref{Fig_conversion}b shows the total transduction efficiency for different pump power combinations with microwave and optical pump powers ranging from 60 to 953\,pW and 92 to 1556\,pW, respectively.
Figure \ref{Fig_conversion}c (\ref{Fig_conversion}d) shows the efficiency versus $P_{\text{o}}\,(P_{\text{e}})$ for fixed microwave (optical) pump power $P_{\text{e}}=601 \,\ (P_{\text{o}}=625)\,\text{pW}$. As expected, the transduction efficiency rises with increasing pump powers and reaches a maximum of $\zeta=1.2\,\%$. The internal transduction efficiency is significantly higher ($\zeta/(\eta_{\text{o}} \eta_{\text{e}})\leq 135\,\%$) because both the microwave resonator as well as the optical cavity are highly under-coupled with coupling ratios of $\eta_{\text{o}}=0.11$ and $\eta_{\text{e}}$ ranging between $0.07-0.18$ when both pumps are on. The increase in the intrinsic loss rate of microwave $\kappa_{\text{in,e}}$ and mechanical resonator $\gamma_{\text{m}}$ at higher pump powers are shown in Fig.~\ref{Fig_conversion}e and Fig.~\ref{Fig_conversion}f caused by considerable heating related to (especially optical) photon absorption. This results in the degradation of the microwave and mechanical quality factors and consequently reduces the waveguide coupling efficiency, the cooperativities and the total conversion efficiency. However, device geometry improvements would increase $\eta_{\text{e}}$ and $\eta_{\text{o}}$ and could improve the total efficiency by up to a factor 100.

In the non-sideband resolved limit 
the contribution of the counter-rotating term of the Hamiltonian $\hat{H}_{\text{CR,o}}$ is non-negligible, resulting in a transduction process that cannot be fully noise-free. This interesting effect can be correctly described by introducing an amplification of the signal tone with (in the absence of thermal noise) quantum limited gain $\mathcal{G}_{\text{o}}$ (see Supplementary Material). In contrast, the microwave resonator is in the resolved-sideband condition $4\omega_{\text{m}}>\kappa_{\text{e}}$, so that the probe tone amplification due to electromechanical interaction is negligible $\mathcal{G}_{\text{e}} \simeq 1$. This results in the total, power independent, bidirectional conversion gain of $\mathcal{G}=\mathcal{G}_{\text{e}}\mathcal{G}_{\text{o}}\simeq\mathcal{G}_{\text{o}}$, 
which turns out to be directly related to the minimum allowed phonon population 
\begin{equation}
\langle n \rangle_{\text{min}}=\frac{(\Delta_{\text{o}}-\omega_{\text{m}})^2+\kappa_{\text{o}}^2/4}{4\Delta_{\text{o}}\omega_{\text{m}}}=\mathcal{G}_{\text{o}}-1
\label{eq_nmin}
\end{equation}
induced by optomechanical quantum backaction when the mechanical resonator is decoupled from its thermal bath~\cite{Aspelmeyer2014}. Due to this amplification process the measured transduction efficiency in Fig.~\ref{Fig_conversion}a is about 110~times larger than one would expect from a model that does not include gain effects for the chosen detuning, and adds the equivalent of at least one vacuum noise photon to the input of the transducer (for $\eta_j=1$ and $\mathcal{G}\gg 1$). However, it turns out that this noise limitation, which might in principle be overcome with active and efficient feedforward \cite{Higginbotham2018},
accounts for only about $0.1\,\%$ of the total conversion noise observed in our system. 
The total transduction (including gain) can be written in terms of the susceptibilities of the electromagnetic modes $\chi_{j}^{-1}(\omega)=i(\Delta_{j}-\omega)+\kappa_{j}/2$ and the mechanical resonator $\chi_{\text{m}}^{-1}(\omega)=i(\omega_{\text{m}}-\omega)+\gamma_{\text{m}}/2$ as
\begin{equation}
\zeta=
\Big|\frac{\sqrt{\kappa_{\text{ex,e}}\kappa_{\text{ex,o}}} G_{\text{e}} G_{\text{o}} \chi_{\text{e}} \chi_{\text{o}}\Big[-\chi_\mathrm{m} + \tilde{\chi}_{\text{m}}\Big]}{1+[\chi_{\text{m}}-\tilde{\chi}_{\text{m}}]\big[G_{\text{e}}^2(\chi_{\text{e}}-\tilde{\chi}_{\text{e}})+G_{\text{o}}^2(\chi_{\text{o}}-\tilde{\chi}_{\text{o}}\big]}\Big|^2,
\label{eq_transduction}
\end{equation}
\\
where $ \tilde{\chi_j}(\omega)=\chi_j(-\omega)^*$. The bandwidth of conversion in our case, where $\kappa_{\text{o}}>4\omega_{\text{m}}$ and $\kappa_{\text{e}}<4\omega_{\text{m}}$, 
is given by $\Gamma_{\text{conv}} \approx (\mathcal{C}_{\text{e}}+1)\gamma_{\text{m}}$.  





\begin{figure}[t]
\centering
\includegraphics[width=1\columnwidth]{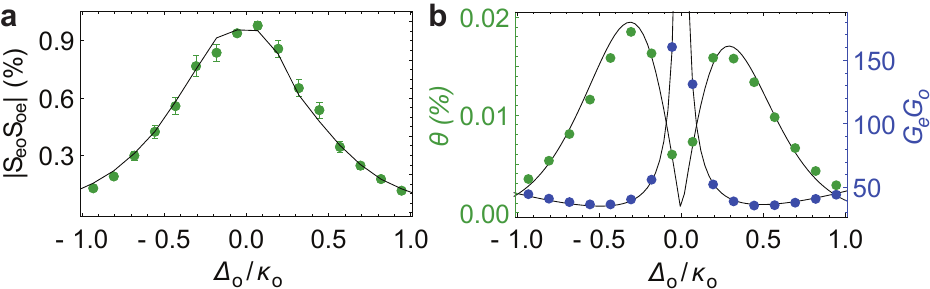}
\caption{\textbf{Pure and amplified conversion.} 
\textbf{a}, Measured total transduction efficiency $\zeta:=|S_{\text{eo}}S_{\text{oe}}|$ with respect to pump detuning $\Delta_\text{o}$ of the non-sideband resolved optical cavity for a constant intracavity photon number $n_{\text{d,o}}=0.185$. 
\textbf{b}, Extracted pure and - in the absence of thermal noise - noiseless microwave-optic conversion $\theta$ as well as the total conversion gain $\mathcal{G}$ that gives rise to amplified vacuum noise. While $\mathcal{G}_{\text{o}}$ diverges for $\Delta_{\text{o}} \rightarrow 0$, the optomechanical damping rate drops to zero 
and leads to a vanishing pure conversion $\theta$.}
\label{Fig_gain}
\end{figure}

As can be seen in Eq.~\ref{eq_nmin}, the signal amplification in the transducer ($\omega \approx \omega_{\text{m}}$) 
depends only on the resonator linewidth and the detuning. As a consequence, it is instructive to measure the transducer parameters as a function of optical pump detuning as shown in Fig.~\ref{Fig_gain}a.
While changing the optical detuning, we also vary the pump power in order to keep the optical intracavity photon number constant at $n_\text{d,o}=0.185 \pm 0.015$. This way it is possible to investigate the influence of $\Delta_{\text{o}}$ at a constant optomechanical coupling $G_{\text{o}}=g_{\text{0,o}}\sqrt{n_{\text{d,o}}}$. The measured total transduction efficiency is shown in Fig.~\ref{Fig_gain}a and reaches $\approx 1\%$ at $\Delta_{\text{o}}=\omega_{\text{m}}$ for the chosen pump powers in agreement with Fig.~\ref{Fig_conversion}c and \ref{Fig_conversion}d. We can now separate the measured transduction (Eq.~\ref{eq_transduction}) into conversion gain and pure conversion $\mathcal{\zeta}:=\mathcal{G}\times \theta$, as shown in Fig.~\ref{Fig_gain}b. The gain shows the expected steep increase at $\Delta_{\text{o}} \rightarrow 0$ where the pure conversion $\theta$ approaches zero
for equal cooling and amplification rates. At $\Delta_{\text{o}}=\kappa_{\text{o}}/2$ on the other hand, where $\langle n \rangle_{\text{min}}$ reaches its minimum of roughly $\kappa_{\text{o}}/4\omega_{\text{m}}\approx 30$, also the gain reaches its minimum and the noiseless part (at zero temperature) of the total conversion process efficiency  shows its highest value of $\theta\approx0.02\%$.

\begin{figure}[t]
\centering
\includegraphics[width=1\columnwidth]{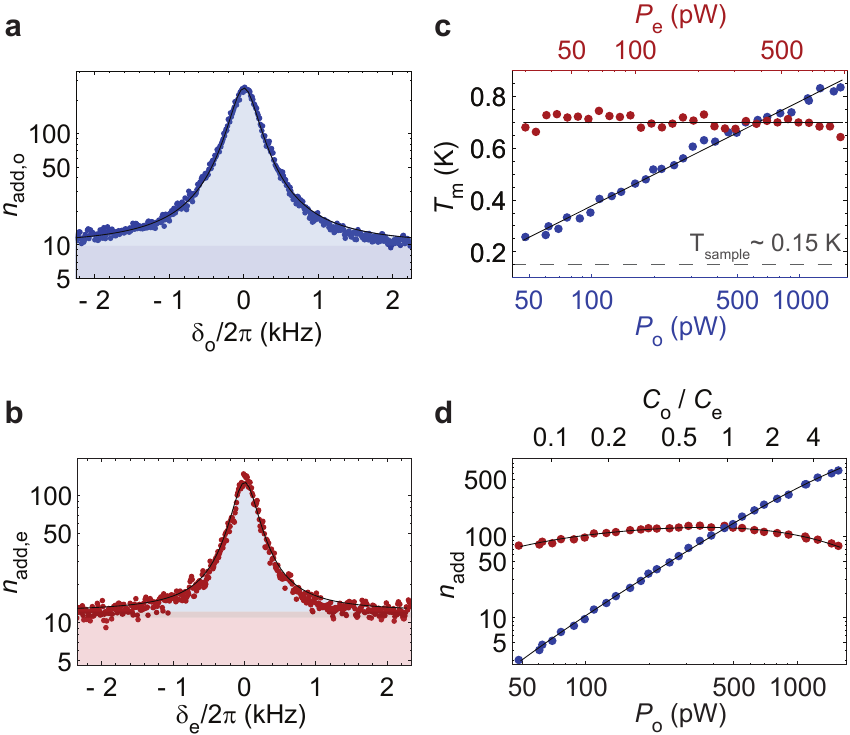}
\caption{\textbf{Conversion noise properties.} Measured noise spectra at the device output for the optical cavity (\textbf{a}) and the microwave resonator (\textbf{b}) as a function of the probe detuning $\delta_j=\omega-(\omega_{\text{d},j}+\omega_{\text{m}})$ at fixed pump powers $P_{\text{e}}=601\,\text{pW}$ and $P_{\text{o}}=625\,\text{pW}$ in units of added noise quanta. \textbf{a}, The dark blue region represents the two-quadrature noise added by the optical measurement chain; the light blue region indicates the thermal mechanical noise added to the converted optical output signal. 
\textbf{b}, 
Bottom light red region represents the two-quadrature background noise from the microwave measurement chain. The central dark red region indicates a small amount of broad band resonator noise and the light blue region the transduction noise due to the thermal population of the mechanical mode. In both panels, fits to one common mechanical bath $n_{\text{m}}$ are shown in black. 
\textbf{c} Mechanical bath temperature $T_\text{m}$ extracted as only fit parameter from fits to the measured output noise as in panels \textbf{a} and \textbf{b} 
with respect to optical (blue dots, $P_{\text{e}}=601\,\text{pW}$) and microwave (red dots, $P_{\text{o}}=625\,\pm\,19\,\text{pW}$) pump power. Black lines show fits to the data with the logarithmic growth function $0.18\,\text{log}_{\text{e}}(P_{\text{o}})-0.47$ and $T_{\text{m}}=0.70\,\text{K}$ respectively. The dashed line indicates the thermalized mechanical mode temperature when the optical pump is off.
\textbf{d}, Microwave (red, $n_{\mathrm{add,e}}$) and optical (blue, $n_{\mathrm{add,o}}$) added noise photons at the output with respect to optical pump power ($P_{\text{e}}=601\,\text{pW}$). The full theory based on an interpolation of $T_{\text{m}}$ from panel~\textbf{c} is shown as black lines.}
\label{Fig_noise}
\end{figure}

Another important figure of merit, not only for quantum applications, is the amount of added noise quanta, usually an effective number referenced to the input of the device. For clarity with regards to the physical origin and the actual measurement of the noise power, in the following we define the total amount of added noise quanta $n_{\text{add},j}$ added to the input signal $S_{\text{in},j}$ after the transduction process as $S_{\text{out},j}=\zeta S_{\text{in},j}+n_{\text{add},j}$. 
Figures~\ref{Fig_noise}a and \ref{Fig_noise}b show the measured conversion noise $n_{\text{add},j}$
as a function of frequency $\delta_j$ at fixed pump powers $P_{\text{e}}=601\,\text{pW}$ and $P_{\text{o}}=625\,\text{pW}$. 
At these powers our device adds $n_{\text{add,o(e)}}=224 (145)$ noise quanta to the output of the microwave resonator (optical cavity), corresponding to an effective input noise of $n_{\text{add,}j}/\zeta$. The noise floor originates from the calibrated measurement system and in case of the microwave port to a small part also from an additional broadband resonator noise, cf.~Fig.~\ref{Fig_noise}b. 
The solid lines are fits to the theory
with the mechanical bath occupation $\bar n_{\text{m}}$ as the only fit parameter (see Supplementary Information).

The fitted effective mechanical bath temperature as a function of pump powers is shown in Fig.~\ref{Fig_noise}c. It reveals the strong optical pump dependent mechanical mode heating (blue), 
while the microwave pump (red) has a negligible influence on the mechanical bath. Figure~\ref{Fig_noise}d shows the measured total added noise at the output of the microwave resonator and optical cavity as a function of optical pump power. The noise added to the optical output (blue) increases with pump power due to absorption heating and increasing optomechanical coupling rate $G_\text{o}$,
while the degradation of the resonator-waveguide coupling efficiency $\eta_{\text{e}}$ explains the decreasing $n_{\mathrm{add,e}}$ at higher optical powers for the microwave output noise (red), see Fig.~\ref{Fig_conversion}e. The intersection of the two noise curves occurs at $\mathcal{C}_{\text{e}}\simeq\mathcal{C}_{\text{o}}$ with cooperativites $C_j$ as defined above and shows that the optical and microwave resonators share the same mechanical thermal bath. The power dependence is in full agreement with theory without free parameters (solid lines) and demonstrates that the thermal mechanical population is the dominating origin of the added transducer noise.

In conclusion, we demonstrated an efficient bidirectional and chip-scale microwave-to-optics transducer using pump powers orders of magnitude lower than comparable all-integrated approaches~\cite{Forsch2020, jiang2019, Fan2018}.
Low pump powers are desired to limit the heat load of the cryostat
and to minimize on-chip heating, which is particularly important for integrated devices because of their limited heat dissipation at millikelvin temperatures. 
Due to the standard material choice involving only silicon and aluminum, our device can be easily integrated with other elements of superconducting circuits and silicon photonic and phononic devices in the future. Comparably simple improvements of its design, especially geometric changes to enhance the waveguide coupling efficiencies,
will significantly increase the total device efficiency. 

While a high conversion performance with minimal pump powers could be demonstrated, optical heating adds incoherent noise to the converted signal due to thermal excitations of the mechanical oscillator. This heating effect can be minimized by better chip thermalization and reduced optical absorption, e.g. using extensive surface cleaning and the reduction of humidity~\cite{Sekoguchi:14}. 
Another possible improvement is to increase the mechanical frequency, which will reduce the number of added noise photons and help to approach the sideband-resolved regime~\cite{Kalaee2019}. Additionally, it has been shown that pulsed pump-probe type experiments can remedy heating effects and, together with high efficiency heralding measurements, be used for post-selecting rare successful conversion or entanglement generation 
events~\cite{Forsch2020,Zhong2020}.
In terms of near-term classical receiver and modulation applications, an important figure of merit is the voltage required to induce an optical phase shift of $\pi$. We are able to reach a value as low as $V_\pi=16\,\mu\text{V}$ (see Supplementary Information), comparable with typical zero point fluctuations in superconducting circuits, nearly a factor 9 lower than the previously reported record \cite{Bagci2014}, and almost $10^{12}$ times more power efficient than commercial passive and wide-band unidirectional electro-optic modulators at X band gigahertz frequencies.


\textbf{Acknowledgements} We thank Yuan Chen for performing supplementary FEM simulations and Andrew Higginbotham, Ralf Riedinger, Sungkun Hong, and Lorenzo Magrini for valuable discussions. This work was supported by IST Austria, the IST nanofabrication facility (NFF), the European Union's Horizon 2020 research and innovation program under grant agreement No 732894 (FET Proactive HOT) and the European Research Council under grant agreement number 758053 (ERC StG QUNNECT). GA is the recipient of a DOC fellowship of the Austrian Academy of Sciences at IST Austria. J.M.F acknowledges support from the Austrian Science Fund (FWF) through BeyondC (F71), a NOMIS foundation research grant, and the EU's Horizon 2020 research and innovation program under grant agreement number 862644 (FET Open QUARTET).
\\
\noindent\textbf{Author contributions}
GA, MW, and SB performed and analyzed the measurements. SB and GA contributed to the theoretical model. GA designed the transducer device. MW, GA, AR and WH built the experimental setup. MW, ER, and GA contributed to sample fabrication. FH tapered optical fibers used for optomechanical characterization tests. GA, MW, SB and JMF wrote the manuscript. JMF supervised the research.
\\
\noindent\textbf{Additional information}
Correspondence and requests for materials should be addressed to JMF.
\\
\textbf{Competing financial interests}
The authors declare no competing financial interests.

\bibliographystyle{naturemag_noURL}
\bibliography{FinkGroupBib_v7}


\newpage
\onecolumngrid
\renewcommand{\thefigure}{S\arabic{figure}}
\appendix
\tableofcontents
\setcounter{figure}{0} 

\section{Theoretical model of the microwave-optical converter} \label{seq_Num_Mod}
\subsection{Hamiltonian}

Our electro-optomechanical system consists of a mechanical resonator with resonance frequency $\omega_{\text{m}}$ that is capacitively coupled to a superconducting microwave resonator and a photonic crystal cavity, as shown in Fig. 1(a) and (b) of the main text. The microwave resonator and optical cavity are driven using microwave and laser pump tones with frequencies $\omega _{\text{d},j}=\omega_{j}-\Delta_{j}$, where $\Delta_{j}$ are the detunings from their resonant frequencies $\omega _{j}$, with $j=\text{e,o}$. We include intrinsic losses for the microwave resonator and optical cavity with rates $\kappa_{\text{in},j}$, and use $\kappa_{\text{ex},j}$ to denote their input-port coupling rates. The Hamiltonian of the coupled system is given by~\cite{Lauk2019}
\begin{eqnarray}
\hat{H} =\hbar \omega_{\text{m}}\hat{b}^{\dagger }\hat{b}+\hbar \sum_{j=\text{e,o}}\Big[\omega_{j}\hat{c}_{j}^{\dagger }\hat{c}_{j}+ g_{0,j}(\hat{b}^{\dagger}+\hat{b})\hat{c}_{j}^{\dagger}\hat{c}_{j}+i E_{j}(\hat{c}_{j}^{\dagger}e^{-i\omega_{\text{d},j}t}-\hat{c}_{j}e^{i\omega_{\text{d},j}t}) \Big],
\end{eqnarray}%
where $\hat{b}$ is the annihilation operator of the mechanical resonator, $\hat{c}_{j}$ is the annihilation operator for
resonator $j=\text{e,o}$ whose coupling rate to the mechanical resonator is $g_{0,j}$. The microwave/optical driving
strength for resonator $j$ is $E_{j}=\sqrt{\kappa_{\text{ex},j}\,P_{j}/\hbar \omega_{\text{d},j}}$, where $P_{j}$ is the power of the driving field~\cite{Barzanjeh2011a}. 

In the interaction picture with respect to $\hbar \sum_{j=\text{e,o}} \omega_{\text{d},j}c_{j}^{\dagger}c_{j}$ and neglecting terms oscillating at $\pm 2\omega_{\text{d},j}$, the system Hamiltonian reduces to
\begin{equation}
\hat{H}=\hbar \omega_{\text{m}}\hat{b}^{\dagger}\hat{b}+\hbar \sum_{j=\text{e,o}}\Big[\Delta_{j}+g_{0,j}(\hat{b}^{\dagger}+\hat{b})\Big]\hat{c}_{j}^{\dagger}\hat{c}_{j}+\hat{H}_{\text{d}}, \label{ham2}
\end{equation}%

where the Hamiltonian associated with the driving fields is $\hat{H}_{\text{d}}=i\hbar \sum_{j=\text{e,o}}E_{j}(\hat{c}_{j}^{\dagger}-\hat{c}_{j})$.

We can linearize Hamiltonian~(\ref{ham2}) by expanding the microwave and optical modes around their steady-state field amplitudes, $\hat{a}_{j}=\hat{c}_{j}-\sqrt{n_{\text{d},j}}$, where $n_{\text{d},j}=|E_{j}|^{2}/(\kappa_{j}^{2}/4+\Delta_{j}^{2})\gg 1$ is the mean number of intracavity photons induced by the microwave and optical pumps~\cite{Barzanjeh2011a}, $\kappa_{j}=\kappa_{\text{in},j}+\kappa_{\text{ex},j}$ are the total resonator decay rates, and $\Delta_{j}$ are the effective resonator and cavity detunings. The linearized Hamiltonian becomes
\begin{equation}
\hat{H}=\hbar \omega_{\text{m}}\hat{b}^{\dagger}\hat{b}+\hbar \sum_{j=\text{e,o}}\Big[\Delta_j\hat{a}_j^{\dagger}\hat{a}_j+G_{j}(\hat{b}+\hat{b}^{\dagger})(\hat{a}_{j}^{\dagger }+\hat{a}_{j})\Big], \label{ham3}
\end{equation}

where $G_{j}=g_{0,j}\sqrt{n_{\text{d},j}}$. By setting the effective resonator detunings so that $\Delta_{\text{e}}=\Delta_{\text{o}}=\omega_{\text{m}}$, moving to an interaction picture, and neglecting the terms rotating at $\pm 2\omega_{\text{m}}$, the above Hamiltonian reduces to
\begin{equation}
\hat{H}=\hbar G_{\text{e}}(\hat{a}_{\text{e}}\hat{b}^{\dagger }+\hat{b}\hat{a}_{\text{e}}^{\dagger })+\hbar G_{\text{o}}(\hat{a}_{\text{o}}\hat{b}^{\dagger }+\hat{b}\hat{a}_{\text{o}}^{\dagger}), \label{hameff}
\end{equation}

as specified in the main text.

\subsection{Equations of motion}
The full quantum treatment of the system can be given in terms of the quantum Langevin equations in which we add to the Heisenberg equations the quantum noise acting on the mechanical resonator ($\hat b_{\text{in}}$ with damping rate $\gamma_m$), as well as the resonator and cavity input fluctuations ($\hat a_{\text{ex},j}$, for $j=\text{e,o}$, with rates $\kappa_{\text{ex},j}$), plus the intrinsic losses of the resonator and cavity modes ($\hat a_{\text{in},j}$, for $j=\text{e,o}$, with loss rates $\kappa_{\text{in},j}$). These noises have the correlation functions
\begin{subequations}
\begin{align}
\langle \hat a_{\text{ext},j}(t) \hat a_{\text{ext},j}^{\dagger}(t^{\prime})\rangle & = \langle \hat a_{\text{ext},j}^{\dagger}(t) \hat a_{\text{ext},j}(t^{\prime})\rangle +\delta(t-t^{\prime})=(\bar{n}_{\text{ext},j}+1)\delta(t-t^{\prime}), \\
\langle \hat a_{\text{in},j}(t) \hat a_{\text{in},j}^{\dagger}(t^{\prime})\rangle & = \langle \hat a_{\text{in},j}^{\dagger}(t) \hat a_{\text{in},j}(t^{\prime})\rangle +\delta(t-t^{\prime})=(\bar{n}_{\text{in},j}+1)\delta(t-t^{\prime}), \\
\langle \hat b_{\text{in}}(t) \hat b_{\text{in}}^{\dagger}(t^{\prime})\rangle & = \langle \hat b_{\text{in}}^{\dagger}(t) \hat b_{\text{in}}(t^{\prime})\rangle +\delta(t-t^{\prime})=(\bar{\text{n}}_m+1)\delta(t-t^{\prime}),
\end{align}

where $\bar{n}_{\text{ext},j}$, $\bar{n}_{\text{in},j}$, and $\bar n_{\text{m}}$ are the Planck-law thermal occupancies of each bath with $j=\text{e,o}$. The resulting Langevin equations corresponding to Hamiltonian~(\ref{ham3}) are 
\end{subequations}
\begin{subequations}
\begin{align} \label{qles2a}
\hat{\dot{a}}_j&=-(\frac{\kappa_{j}}{2}+i\Delta_j) \hat a_{j}-iG_{j}(\hat b+\hat b^\dagger)+\sqrt{\kappa_{\text{ext},j}}\hat a_{
\text{ex},j}+\sqrt{\kappa_{\text{in},j}}\hat a_{\text{in},j},\\
\hat{\dot{b}}&=-(\frac{\gamma_{\text{m}}}{2}+i\omega_{\text{m}}) \hat b-i\sum_{j=\text{e,o}}G_{{j}}(\hat a_{{j}}+\hat a_{{j}}^{\dagger})+\sqrt{\gamma_{\text{m}}}\hat b_{\text{in}}.\label{qles2c}
\end{align}
\end{subequations}

We can solve the above equations in the Fourier domain to obtain the microwave resonator and optical cavity variables. By substituting the solutions of Eqs.~(\ref{qles2a})--(\ref{qles2c}) into the corresponding input-output relation, i.e., $\hat a_{\text{out},j}= \sqrt{\kappa_{\text{ex},j}}\hat a_j-\hat a_{\text{ex},j}$, we obtain
\begin{equation}\label{equationmatrix}
\mathbf{S}_{\text{out}}(\omega)=\mathbf{\Upsilon}(\omega)\mathbf{S}_{\text{in}}(\omega),
\end{equation}

where $\mathbf{\Upsilon}(\omega)=\Big(\mathbf{C}.[-i\omega \mathbf{I}-\mathbf{A}]^{-1}.\mathbf{B}-\mathbf{D}\Big)$ with $\textbf{I}$ is the identity matrix, $\textbf{S}_{\text{out}}=[\hat{a}_{\text{out,e}},\hat{a}_{\text{out,o}},\hat{a}_{\text{out,e}}^{\dagger},\hat{a}_{\text{out,o}}^{\dagger}]^T$, $\textbf{S}_{\text{in}}=[\hat{a}_{\text{ext,e}},\hat{a}_{\text{in,e}},\hat{a}_{\text{ext,o}},\hat{a}_{\text{in,o}},\hat{b}_{\text{in}},\hat{a}_{\text{ext,e}}^{\dagger},\hat{a}_{\text{in,e}}^{\dagger},\hat{a}_{\text{ext,o}}^{\dagger},\hat{a}_{\text{in,o}}^{\dagger},\hat{b}_{\text{in}}^{\dagger}]^T$, and we have defined the following matrices
\begin{equation}
\mathbf{A}=\begin{bmatrix}
-(\frac{\kappa_{\text{e}}}{2}+i\Delta_{\text{e}}) & 0 & -iG_{\text{e}} & 0 & 0 & -iG_{\text{e}}\\
0 & -(\frac{\kappa_{\text{o}}}{2}+i\Delta_{\text{o}})  & -iG_{\text{o}} & 0 & 0 & -iG_{\text{o}} \\
-iG_{\text{e}} & -iG_{\text{o}} & -(\frac{\gamma_{\text{m}}}{2}+i\omega_{\text{m}}) & -iG_{\text{e}} & -iG_{\text{o}} & 0 \\
0 & 0 & iG_{\text{e}} & -(\frac{\kappa_{\text{e}}}{2}-i\Delta_{\text{e}}) & 0 & iG_{\text{e}}\\
0 & 0 & iG_{\text{o}} & 0 & -(\frac{\kappa_{\text{o}}}{2}-i\Delta_{\text{o}}) & iG_{\text{o}}\\ 
iG_{\text{e}} & iG_{\text{o}} & 0 & iG_{\text{e}} & iG_{\text{o}}  & -(\frac{\gamma_{\text{m}}}{2}-i\omega_{\text{m}})
\end{bmatrix},
\end{equation}

\begin{equation}
\mathbf{B}=\begin{bmatrix}
\sqrt{\kappa_{\text{e}}\eta_{\text{e}}} & \sqrt{\kappa_{\text{e}}(1-\eta_{\text{e}})} & 0 & 0 &0 & 0 &0 & 0 &0 & 0 \\
0 & 0 & \sqrt{\kappa_{\text{o}}\eta_{\text{o}}} & \sqrt{\kappa_{\text{o}}(1-\eta_{\text{o}})} &0 & 0 &0 & 0 &0 & 0 \\
0 & 0 & 0 & 0 &\sqrt{\gamma_{\text{m}}}  & 0 &0 & 0 &0 & 0 \\
0 & 0 & 0 & 0 &0 &\sqrt{\kappa_{\text{e}}\eta_{\text{e}}} & \sqrt{\kappa_{\text{e}}(1-\eta_{\text{e}})} & 0 &0 & 0 \\
0 & 0 & 0 & 0 &0 & 0 &0 & \sqrt{\kappa_{\text{o}}\eta_{\text{o}}} & \sqrt{\kappa_{\text{o}}(1-\eta_{\text{o}})} & 0 \\
0 & 0 & 0 & 0 &0 & 0 &0 & 0 &0 & \sqrt{\gamma_{\text{m}}} 
\end{bmatrix},
\end{equation}

\begin{equation}
\mathbf{C}=\begin{bmatrix}
\sqrt{\kappa_{\text{e}}\eta_{\text{e}}} &0 & 0 & 0 &0 & 0 \\
0 & \sqrt{\kappa_{\text{o}}\eta_{\text{o}}} & 0 & 0 &0 & 0 \\
0 & 0 & 0 & \sqrt{\kappa_{\text{e}}\eta_{\text{e}}} &0 &0 \\
0 & 0 & 0 & 0 &\sqrt{\kappa_{\text{o}}\eta_{\text{o}}} & 0
\end{bmatrix},
\end{equation}

\begin{equation}
\mathbf{D}=\begin{bmatrix}
1 & 0 & 0 & 0 &0 & 0 &0 & 0 &0 & 0 \\
0 & 0 & 1 & 0 &0 & 0 &0 & 0 &0 & 0 \\
0 & 0 & 0 & 0 &0 & 1 &0 & 0 &0 & 0 \\
0 & 0 & 0 & 0 &0 &0 & 0 & 1 &0 & 0 
\end{bmatrix},
\end{equation}
here $\eta_j=\kappa_{\text{ext,j}}/\kappa_j$. 

The total output fields are then
\begin{subequations} \label{coeff0}
\begin{gather}
\hat a_\text{out,e}=(\eta_{\text{e}}\alpha_{\text{e,e}}-1)\hat a_\text{ext,e}+\sqrt{\eta_{\text{e}}}\Bigl[\sqrt{1-\eta_{\text{e}}}\alpha_{\text{e,e}}\hat a_\text{int,e}+\sqrt{\eta_{\text{o}}}\alpha_{\text{e,o}}\hat a_\text{ext,o}+\sqrt{1-\eta_{\text{o}}}\alpha_{\mathrm{e,o}}\hat a_\text{int,o}+\alpha_{\text{e,m}}\hat b_\text{in}
\nonumber\\
+\sqrt{\eta_{\text{e}}}\tilde \alpha_{\text{e,e}}\hat a_\text{ext,e}^{\dagger}+\sqrt{1-\eta_{\text{e}}}\tilde \alpha_{\text{e,e}}\hat a_\text{int,e}^{\dagger}+\sqrt{\eta_{\text{o}}}\tilde \alpha_{\text{e,o}}\hat a_\text{ext,o}^{\dagger}+\sqrt{1-\eta_{\text{o}}}\tilde \alpha_{\text{e,o}}\hat a_\text{int,o}^{\dagger}+\tilde \alpha_{\text{e,m}}\hat b_\text{in}^{\dagger}\Bigr],
\\
\hat a_\text{out,o}=(\eta_{\text{o}}\alpha_{\text{o,o}}-1)\hat a_\text{ext,o}+\sqrt{\eta_{\text{o}}}\Bigl[\sqrt{1-\eta_{\text{o}}}\alpha_{\text{o,o}}\hat a_\text{int,o}+\sqrt{\eta_{\text{e}}}\alpha_{\text{o,e}}\hat a_\text{ext,e}+\sqrt{1-\eta_{\text{e}}}\alpha_{\text{o,e}}\hat a_\text{int,e}+\alpha_{\text{o,m}}\hat b_\mathrm{in}
\nonumber\\
+\sqrt{\eta_{\text{o}}}\tilde \alpha_{\text{o,o}}\hat a_\text{ext,o}^{\dagger}+\sqrt{1-\eta_{\text{o}}}\tilde \alpha_{\text{o,o}}\hat a_\text{int,o}^{\dagger}+\sqrt{\eta_{\text{e}}}\tilde \alpha_{\text{o,e}}\hat a_\text{ext,e}^{\dagger}+\sqrt{1-\eta_{\text{e}}}\tilde \alpha_{\text{o,e}}\hat a_\text{int,e}^{\dagger}+\tilde \alpha_{\text{o,m}}\hat b_\text{in}^{\dagger}\Bigr],
\end{gather}
\end{subequations}

with $\hat a_{\text{ext},j}^{({\dagger})}$ and $\hat a_{\text{int},j}^{({\dagger})}$ referring to modes in the waveguide and bath respectively and the coefficients
\begin{subequations}\label{coeff}
\begin{align}  
\alpha_{\text{e,e}}&=\frac{\kappa_\text{e}\, \chi_\text{e}\Big(1+G_\text{o}^2(\chi_\text{o}-\chi_\text{o}^*)-G_\text{e}^2\chi_\text{e}^*\Big)\Big[-\chi_\text{m} + \chi_\text{m}^*\Big]}{1+[\chi_{\text{m}}-\chi_{\text{m}}^*]\big[G_{\text{e}}^2(\chi_{\text{e}}-\chi_{\text{e}}^*)+G_{\text{o}}^2(\chi_{\text{o}}-\chi_{\text{o}}^*)\big]},\\
\alpha_{\text{o,o}}&=\frac{\kappa_\text{o}\, \chi_\text{o}\Big(1+G_\text{e}^2(\chi_\text{e}-\chi_\text{e}^*)-G_\text{o}^2\chi_\text{o}^*\Big)\Big[-\chi_{\text{m}} + \chi_{\text{m}}^*\Big]}{1+[\chi_{\text{m}}-\chi_{\text{m}}^*]\big[G_{\text{e}}^2(\chi_{\text{e}}-\chi_{\text{e}}^*)+G_{\text{o}}^2(\chi_{\text{o}}-\chi_{\text{o}}^*)\big]},\\
\alpha_{\text{e,o}}&=\alpha_{\text{o,e}}=\frac{\sqrt{\kappa_\text{e}\kappa_\text{o}}\, \chi_{\text{e}}\chi_{\text{o}}G_{\text{e}}G_{\text{o}}\Big[-\chi_{\text{m}} + \chi_{\text{m}}^*\Big]}{1+[\chi_{\text{m}}-\chi_{\text{m}}^*]\big[G_{\text{e}}^2(\chi_{\text{e}}-\chi_{\text{e}}^*)+G_{\text{o}}^2(\chi_{\text{o}}-\chi_{\text{o}}^*)\big]},\\
\alpha_{j,\text{m}}&=-\frac{i\,\sqrt{\kappa_j\gamma_\text{m}}\, G_j\chi_j\chi_\text{m}}{1+[\chi_{\text{m}}-\chi_{\text{m}}^*]\big[G_{\text{e}}^2(\chi_{\text{e}}-\chi_{\text{e}}^*)+G_{\text{o}}^2(\chi_{\text{o}}-\chi_{\text{o}}^*)\big]},\\
\tilde \alpha_{j,j}&=\frac{\kappa_j\, \chi_j \chi_j^* G_j^2\Big[-\chi_{\text{m}} + \chi_{\text{m}}^*\Big]}{1+[\chi_{\text{m}}-\chi_{\text{m}}^*]\big[G_{\text{e}}^2(\chi_{\text{e}}-\chi_{\text{e}}^*)+G_{\text{o}}^2(\chi_{\text{o}}-\chi_{\text{o}}^*)\big]},\\
\tilde \alpha_\text{e,o}&=\frac{\sqrt{\kappa_\text{e}\kappa_\text{o}}\, \chi_\text{e}\chi_\text{o}^*G_\text{e}G_\text{o}\Big[-\chi_{\text{m}} + \chi_{\text{m}}^*\Big]}{1+[\chi_{\text{m}}-\chi_{\text{m}}^*]\big[G_{\text{e}}^2(\chi_{\text{e}}-\chi_{\text{e}}^*)+G_{\text{o}}^2(\chi_{\text{o}}-\chi_{\text{o}}^*)\big]},\\
\tilde \alpha_\text{o,e}&=-\frac{\sqrt{\kappa_\text{e}\kappa_\text{o}}\, \chi_\text{e}^*\chi_\text{o}G_\text{e}G_\text{o}\Big[-\chi_{\text{m}} + \chi_{\text{m}}^*\Big]}{1+[\chi_{\text{m}}-\chi_{\text{m}}^*]\big[G_{\text{e}}^2(\chi_{\text{e}}-\chi_{\text{e}}^*)+G_{\text{o}}^2(\chi_{\text{o}}-\chi_{\text{o}}^*)\big]},\\
\tilde \alpha_{j,\text{m}}&=-\frac{i\,\sqrt{\kappa_j\gamma_\text{m}}\, G_j\chi_j\chi_\text{m}^*}{1+[\chi_{\text{m}}-\chi_{\text{m}}^*]\big[G_{\text{e}}^2(\chi_{\text{e}}-\chi_{\text{e}}^*)+G_{\text{o}}^2(\chi_{\text{o}}-\chi_{\text{o}}^*)\big]},
\end{align}
\end{subequations}

with $j=\text{e,o}$. We also define the individual susceptibilities of the optical cavity and microwave resonator $\chi_{j}^{-1}=\chi_{j}(\omega)^{-1}=i(\Delta_{j}-\omega)+\kappa_{j}/2$ and the mechanical susceptibility $\chi_{\text{m}}^{-1}=\chi_{\text{m}}(\omega)^{-1}=i(\omega_{\text{m}}-\omega)+\gamma_{\text{m}}/2$ and $\chi_{k}^*=\chi_{k}(-\omega)^*$ $k=\text{e,o,m}$.

Note that the commutation relation $[\hat a_{\text{out},j}(\omega),\hat a_{\text{out},j}^\dagger(\omega')]=\delta(\omega-\omega')$ imposes the following constrains
\begin{subequations}
\begin{align}
(|\eta_{\text{e}}\alpha_\text{e,e}-1|^2+\eta_{\text{e}}(1-\eta_{\text{e}})|\alpha_\text{e,e}|^2-\eta_{\text{e}}|\tilde \alpha_\text{e,e}|^2)+\eta_{\text{e}}(|\alpha_\text{e,o}|^2-|\tilde \alpha_\text{e,o}|^2)+\eta_{\text{e}}(|\alpha_\text{e,m}|^2-|\tilde \alpha_\text{e,m}|^2)&=1,\\
(|\eta_{\text{o}}\alpha_\text{o,o}-1|^2+\eta_{\text{o}}(1-\eta_{\text{o}})|\alpha_\text{o,o}|^2-\eta_{\text{o}}|\tilde \alpha_\text{o,o}|^2)+\eta_{\text{o}}(|\alpha_\text{o,e}|^2-|\tilde \alpha_\text{o,e}|^2)+\eta_{\text{o}}(|\alpha_\text{o,m}|^2-|\tilde \alpha_\text{o,m}|^2)&=1.
\end{align}
\end{subequations}

\subsection{Conversion efficiency and gain}

From equation (\ref{equationmatrix}) we can directly calculate all elements of the scattering matrix including the reflection parameters and transduction efficiency. The \textit{effective} microwave-to-optical transduction efficiency is given by 
 \begin{equation}\label{conversion0}
 \zeta(\omega):=|\Upsilon_{2,1}|^2=|\Upsilon_{1,2}|^2=\Big|\frac{\sqrt{\kappa_{\text{ex,e}}\kappa_\text{ex,o}} G_\text{e} G_\text{o} \chi_\text{e}(\omega) \chi_\text{o}(\omega)\Big[\mathrm{-\chi_\text{m}(\omega) + \chi_\text{m}(-\omega)^*}\Big]}{1+[\chi_\text{m}(\omega)-\chi_\text{m}(-\omega)^*]\big[G_\text{e}^2(\chi_\text{e}(\omega)-\chi_\text{e}(-\omega)^*)+G_\text{o}^2(\chi_\text{o}(\omega)-\chi_\text{o}(-\omega)^*)\big]}\Big|^2.
\end{equation}

The above equation contains the pure conversion efficiency and the gain due to the unresolved condition of the optical mode. We can separate these two effects by rewriting Eq.~(\ref{conversion0}) in terms of the electro- and optomechanical damping rates $\Gamma_{j}=G^2_{j}\Big[\frac{\kappa_j}{(\Delta_j-\omega)^2+\kappa_j^2/4}-\frac{\kappa_j}{(\Delta_j+\omega)^2+\kappa_j^2/4}\Big]$. This then gives $ \zeta(\omega)=\theta\times \mathcal{G}$, where
 \begin{equation}\label{conversion01}
 \theta=\Big|\frac{2\sqrt{\eta_{\text{e}}\eta_\text{o}}\,\sqrt{\Gamma_\text{e}\Gamma_\text{o}}}{2i(\omega-\omega_\text{m}')+\gamma_{\text{m}}+\Gamma_{\text{e}}+\Gamma_{\text{o}}}\Big|^2,
\end{equation}

is the pure bidirectional optical-to-microwave conversion efficiency with $\omega_\text{m}'=\omega_\text{m}-\delta_\omega (\omega_\text{m})$ and $\delta_\omega (\omega_\text{m})=\sum_{j=\text{e,o}}\text{Im} (G_j^2(\chi_j-\chi_j^*))$ being the electro- and optomechanical frequency shifts while $ \mathcal{G}= \mathcal{G}_\text{o} \mathcal{G}_\text{e}$ is the amplification gain of the converter where
\begin{subequations}\label{conversion001}
\begin{align}
 \mathcal{G}_\text{e}&=\Big(\frac{|\chi_\text{e}|^2}{4\Delta_\text{e}\omega_\text{m}}\Big)\Big[(\Delta_\text{e}-\omega)^2+\kappa_\text{e}^2/4\Big]\Big[(\Delta_\text{e}+\omega)^2+\kappa_\text{e}^2/4\Big],\\
 \mathcal{G}_\text{o}&=\Big(\frac{|\chi_\text{o}|^2}{4\Delta_\text{o}\omega_\text{m}}\Big)\Big[(\Delta_\text{o}-\omega)^2+\kappa_\text{o}^2/4\Big]\Big[(\Delta_\text{o}+\omega)^2+\kappa_\text{o}^2/4\Big],
\end{align}
\end{subequations}

are the gains attributed to the unresolved sideband condition of the optical cavity and the microwave resonator. 
For our system $\delta_\omega \ll \omega_\text{m}$, as such we consider $\omega=\omega_\text{m}'\simeq \omega_\text{m}$, resulting in
\begin{subequations}
\begin{align}
  \mathcal{G}_\text{e}&=\Big(\frac{(\Delta_\text{e}+\omega_\text{m})^2+\kappa_\text{e}^2/4}{4\Delta_\text{e}\, \omega_\text{m}}\Big),\\  
  \mathcal{G}_\text{o}&=\Big(\frac{(\Delta_\text{o}+\omega_\text{m})^2+\kappa_\text{o}^2/4}{4\Delta_\text{o}\, \omega_\text{m}}\Big),
\end{align}
\end{subequations}

Note that $\Delta_\text{e}=\omega_\text{m}$ and considering the fact that in our system the microwave resonator is in the resolved sideband regime $\omega_\text{m}\gg \kappa_\text{e}$ entails $\mathcal{G}_\text{e}\simeq 1$. The total gain, therefore, reduces to $\mathcal{G}\simeq\mathcal{G}_{\text{o}}=1+\langle n \rangle_{\text{min}}$ where
\begin{equation}
\langle n \rangle_{\text{min}}=\frac{(\Delta_\text{o}-\omega_\text{m})^2+\kappa_\text{o}^2/4}{4\Delta_\text{o}\omega_\text{m}},
\end{equation}

is the minimum phonon number of the mechanical resonator induced by the optomechanical quantum backaction when the mechanical resonator is decoupled from its thermal bath~\cite{Marquardt2007, Wilson-Rae2007}. At the optical detuning $\Delta_{\text{o}}=\kappa_\text{o}/2$ the phononic occupation number at absence of thermal noise reaches its minimum $\langle n \rangle_{\text{min}} \simeq \kappa_{\text{o}}/4\omega_{\text{m}}\gg 1$. In this regime the backaction cooling of the mechanical resonator to its ground state is prohibited.\\

We can rewrite Eqs.~(\ref{coeff}) in terms of the system gains
\begin{equation}
\begin{aligned}
\eta_{\text{e}}\eta_{\text{o}} |\tilde \alpha_\text{e,o}|^2 &= \theta \mathcal{G}_\text{e}(\mathcal{G}_\text{o}-1),\\
\eta_{\text{e}}\eta_{\text{o}} |\tilde \alpha_\text{o,e}|^2 &= \theta \mathcal{G}_\text{o}(\mathcal{G}_\text{e}-1),\\
\eta_{\text{e}}\eta_{\text{o}} |\alpha_\text{e,o}|^2 &= \eta_{\text{e}}\eta_{\text{o}} |\alpha_\text{o,e}|^2 = \theta \mathcal{G}_\text{o} \mathcal{G}_\text{e}.
\end{aligned}
\end{equation}
Using the above equation we can simplify Eqs.~(\ref{coeff0}) to
\begin{subequations}\label{equationAmp1}
\begin{align} 
\hat a_\mathrm{out,e}/\sqrt{\theta}=\mathcal{G}_\mathrm{e}(\sqrt{\mathcal{G}_\mathrm{o}}\hat a_\mathrm{ext,o}+\sqrt{\mathcal{G}_\mathrm{o}-1}\hat a_\mathrm{ext,o}^{\dagger})+\sum_{j=e,o}\sum_{i=e,o,m}F_\mathrm{e}(\alpha_{j,i}/\sqrt{\theta},\hat{O}),\\
\hat a_\mathrm{out,o}/\sqrt{\theta}=\mathcal{G}_\mathrm{o}(\sqrt{\mathcal{G}_\mathrm{e}}\hat a_\mathrm{ext,e}+\sqrt{\mathcal{G}_\mathrm{e}-1}\hat a_\mathrm{ext,e}^{\dagger})+\sum_{j=e,o}\sum_{i=e,o,m}F_\mathrm{o}(\alpha_{j,i}/\sqrt{\theta},\hat{O}),
\end{align}
\end{subequations}
The terms inside the brackets in the RHS of the above equations describe the amplification of the quantum fluctuation $\hat a_\mathrm{ext,o (e)}$ at the input port of the optical cavity (microwave resonator) with corresponding gain $\mathcal{G}_\mathrm{o(e)}$ \cite{caves_quantum_1982}. Here, $F_\mathrm{e (o)}(\alpha_{i,j}/\sqrt{\theta},\hat{O})$ show the contribution of the quantum fluctuation at the input of the microwave resonator (optical cavity) and mechanical resonator. 

In the resonance condition $\omega=\Delta_j=\omega_\text{m}$, the total gain simplifies to $ \mathcal{G}=\mathcal{G}_\text{o} \mathcal{G}_\text{e}=[1+(\kappa_\text{o}/4\omega_\text{m})^2][1+(\kappa_\text{e}/4\omega_\text{m})^2]$.\\
If electro- and optomechanical cavity are additionally in the resolved sideband regime ($\omega_\text{m}\gg \kappa_j$) all contributions from counter-rotating terms in the Hamiltonian (\ref{ham3}) become negligible, resulting in $\mathcal{G}=1$ and the \textit{effective} conversion efficiency (\ref{conversion0}) reduces to
\begin{equation}\label{conversion1}
\mathrm{\zeta_{sbr}(\omega_\text{m})=\frac{4 \eta_\text{e}\eta_\text{o} \Gamma_\text{e} \Gamma_\text{o}} {(\gamma_\text{m} + \Gamma_\text{e} + \Gamma_\text{o})^2}=\frac{4 \eta_\text{e}\eta_\text{o}\mathcal{C}_\text{e} \mathcal{C}_\text{o}} {(1 + \mathcal{C}_\text{e} + \mathcal{C}_\text{o})^2}}.
\end{equation}

where $\Gamma_j$ simplifies to $ \Gamma_{j} = \frac{4 g_{0,j}^2 n_{\text{d},j}} {\kappa_{j}} = \frac{4G_j^2}{\kappa_j} = \mathcal{C}_j \gamma_\text{m}$ with the optomechanical cooperativity $\mathcal{C}_j$.
\\
\\
\subsection{Conversion bandwidth} 
The bandwidth of the conversion process can be calculated from the denominator of equation (\ref{conversion0}). In our experiment the microwave resonator is in the resolved sideband regime $4\omega_\text{m}\gg \kappa_\text{e}$, while the optical cavity goes beyond this regime $\omega_\text{m}\ll \kappa_\text{o}$. As such for $\omega=\Delta_j=\omega_\text{m}\gg \gamma_\text{m}$ we have $\chi_\text{e}\rightarrow  2/\kappa_\text{e} $, $\chi_\text{o}\simeq \chi_\text{o}^*\rightarrow 2/\kappa_\text{o}$, $\chi_\text{m}\rightarrow 2/\gamma_\text{m}$, and $\{\chi_\text{e}^*,\chi_\text{m}^*\}\rightarrow 0 $ which gives the following bandwidth 
\begin{equation}
\label{eq_bandwidth00}
\Gamma_\text{conv} \approx \Gamma_\text{e}+\gamma_\text{m},
\end{equation}

Its dependence on the optomechanical damping rate $\Gamma_\text{o}\ll \Gamma_\text{e}$ is negligible because the unresolved condition of the optical cavity significantly degrades $\Gamma_\text{o}$ due to equal photon scattering to the red and blue sidebands. 
\\
\\
\subsection{Added noise} 
The total noise added during conversion including the vibrational noise of the mechanics and the resonators' noises can be calculated with the spectral density of the output fields
\begin{equation}
\label{eq_total_noise}
2\pi\, \mathbf{ S}_{\text{SD}}(\omega)\delta(\omega-\omega')=\langle \mathbf{ S}_{\text{out}}(\omega')^\dagger \mathbf{ S}_{\text{out}}(\omega)\rangle.
\end{equation}

Using Eq.~(\ref{coeff0}), the total noises added to the output of the microwave resonator and optical cavity are given by
\begin{subequations}
\label{noise}
\begin{multline}  
n_\text{add,e} = |\eta_{\text{e}}\alpha_{\text{e,e}}-1|^2 \bar n_\text{ext,e}
+\eta_{\text{e}}\Bigl[
(1-\eta_{\text{e}})|\alpha_{\text{e,e}}|^2 \bar n_\text{int,e}
+\eta_{\text{o}}|\alpha_{\text{e,o}}|^2 \bar n_\text{ext,o}
+(1-\eta_{\text{o}})|\alpha_{\text{e,o}}|^2 \bar n_\text{int,o}
+|\alpha_{\text{e,m}}|^2 \bar n_\text{m}
\\
+\eta_{\text{e}}|\tilde \alpha_{\text{e,e}}|^2(\bar n_\text{ext,e}+1)
+(1-\eta_{\text{e}})|\tilde \alpha_{\text{e,e}}|^2(\bar n_\text{int,e}+1)
+\eta_{\text{o}}|\tilde \alpha_{\text{e,o}}|^2(\bar n_\text{ext,o}+1)
+(1-\eta_{\text{o}})|\tilde \alpha_{\text{e,o}}|^2(\bar n_\text{int,o}+1)
+|\tilde\alpha_{\text{e,m}}|^2(\bar n_\text{m}+1)\Bigr],
\end{multline}
\begin{multline}
n_\text{add,o}=|\eta_{\text{o}}\alpha_{\text{o,o}}-1|^2 \bar n_\text{ext,o}
+\eta_{\text{o}}\Bigl[
(1-\eta_{\text{o}})|\alpha_{\text{o,o}}|^2 \bar n_\text{int,o}
+\eta_{\text{e}}|\alpha_{\text{o,e}}|^2 \bar n_\text{ext,e}
+(1-\eta_{\text{e}})|\alpha_{\text{o,e}}|^2 \bar n_\text{int,e}
+|\alpha_{\text{o,m}}|^2 \bar n_\text{m}
\\
+\eta_{\text{o}}|\tilde \alpha_{\text{o,o}}|^2(\bar n_\text{ext,o}+1)
+(1-\eta_{\text{o}})|\tilde \alpha_{\text{o,o}}|^2(\bar n_\text{int,o}+1)
+\eta_{\text{e}}|\tilde \alpha_{\text{o,e}}|^2(\bar n_\text{ext,e}+1)
+(1-\eta_{\text{e}})|\tilde \alpha_{\text{o,e}}|^2(\bar n_\text{int,e}+1)
+|\tilde\alpha_{\text{o,m}}|^2(\bar n_\text{m}+1)\Bigr].
\end{multline}
\end{subequations}

The noise terms can be simplified in the vacuum condition in which the thermal occupations of the microwave resonator $\bar n_\text{ext,e}=\bar n_\text{int,e}=0$, optical cavity $\bar n_\text{ext,o}=\bar n_\text{int,o}=0$, and mechanical resonator $\bar n_\text{m}=0$ are negligible,
\begin{subequations} \label{noise}
\begin{align} 
n_\text{add,e}&=\eta_{\text{e}}\Bigl(|\tilde\alpha_{\text{e,e}}|^2
+|\tilde\alpha_{\text{e,o}}|^2
+|\tilde\alpha_{\text{e,m}}|^2 \Bigr),\\
n_\text{add,o}&=\eta_{\text{o}}\Bigl(|\tilde\alpha_{\text{o,o}}|^2
+|\tilde\alpha_{\text{o,e}}|^2
+|\tilde\alpha_{\text{o,m}}|^2 \Bigr).
\end{align}
\end{subequations}

We can write the noise added to the output of the transducer in terms of the electromechanical and optomechanical gain introduced in Eqs.~(\ref{conversion001}). By considering $|\tilde\alpha_{\text{e(o),m}}|_{\omega=\omega_{\text{m}}}^2\ll \{|\tilde\alpha_{\text{e,o}}|^2,|\tilde\alpha_{\text{e,o}}|^2,|\tilde\alpha_{j,j}|^2\}$ (since  $\chi_\text{m}(\omega_\text{m}) \gg \chi_\text{m}^*(-\omega_\text{m})$ for $\omega_\text{m} \gg \gamma_\text{m}$) the Eqs. (\ref{noise}) reduce to 
\begin{subequations} \label{noiseall}
\begin{align} \label{noise44}
n_\text{add,e}&\simeq \theta/\eta_{\text{o}}\, \mathcal{G}_{\text{e}} \Big[\Big(\frac{\Gamma_{\text{e}}}{\Gamma_{\text{o}}}\Big)(\mathcal{G}_{\text{e}}-1)+(\mathcal{G}_{\text{o}}-1)\Big],\\
n_\text{add,o}&\simeq\theta/\eta_{\text{e}}\, \mathcal{G}_{\text{o}} \Big[\Big(\frac{\Gamma_{\text{o}}}{\Gamma_{\text{e}}}\Big)(\mathcal{G}_{\text{o}}-1)+(\mathcal{G}_{\text{e}}-1)\Big].
\end{align}\label{noise55}
\end{subequations}

The above equations can be simplified further and written in terms of the added noises of the optical and microwave amplifier models introduced in Eqs.~(\ref{equationAmp1}). The resolved sideband condition of either the microwave resonator or optical cavity, respectively, results in $\mathcal{G}_{\mathrm{e}}\simeq 1$ or $\mathcal{G}_{\mathrm{o}}\simeq 1$, as such Eqs.~(\ref{noise55}) reduce to
\begin{subequations}\label{noiseSimplify11}
\begin{align}
n_{\mathrm{amp,e}}&=\frac{n_\mathrm{add,e}}{\theta}\simeq \mathcal{G}_{\mathrm{o}}-1,\nonumber\\
n_{\mathrm{amp,o}}&=\frac{n_\mathrm{add,o}}{\theta}\simeq \mathcal{G}_{\mathrm{e}}-1,
\end{align}
\end{subequations}
representing the added noise at the output of quantum limited amplifiers with gains $\mathcal{G}_{\mathrm{o}}$ and $\mathcal{G}_{\mathrm{e}}$ considering vacuum noise at the input ports, in agreement with Eqs. (\ref{equationAmp1}).

For the special case of $\Delta_\text{e}=\omega_{\text{m}}$ and considering the microwave resonator being in the resolved sideband condition i.e. $\mathcal{G}_{\text{e}}\simeq 1$, we can rewrite Eqs. (\ref{noiseall}) in terms of the phononic occupancy $\langle n \rangle_{\text{min}}$, as 
\begin{subequations}\label{noiseadd}
\begin{align}
n_\text{add,e}&\simeq\theta/\eta_{\text{o}}\,(\mathcal{G}_{\text{o}}-1)=\theta/\eta_{\text{o}}\,\langle n \rangle_{\text{min}},\\
n_\text{add,o}&\simeq\theta/\eta_{\text{e}}\,(\mathcal{G}_{\text{o}}-1)\mathcal{G}_{\text{o}}\Big(\frac{\Gamma_{\text{o}}}{\Gamma_{\text{e}}}\Big)=\theta/\eta_{\text{e}}\, \langle n \rangle_{\text{min}} \big(\langle n \rangle_{\text{min}}+1\big) \Big(\frac{\Gamma_{\text{o}}}{\Gamma_{\text{e}}}\Big).
\end{align}
\end{subequations}

Note that for $\theta=1$ and therefore also $\eta_j=1$, the added noises in Eq. (\ref{noiseadd})a represents the amplification of the vacuum noise with gain $\mathcal{G}_\text{o}$ which is the direct result of the quantum backaction induced phononic occupation $\langle n \rangle_{\text{min}} = \mathcal{G}_\text{o}-1$.



\section{Performance as a classical phase modulator}

Considering our device as a classical phase modulator the most important figure of merit is its value for $V_\pi$. This value represents the voltage amplitude of an input microwave signal that generates a $\pi$ phase shift of the optical output with respect to its input.
In this section we derive the equation for calculating $V_\pi$ for a triple-resonant electro-opto-mechanical system. Moreover, we also estimate the required energy $E_\text{bit}$ to encode a classical bit on an optical carrier signal. Finally, we compare the values of both figures of merit to the literature.

\subsection{Modulation voltage $V_\pi$}
Starting point for the derivation of $V_\pi$ is the relation between the optomechanical coupling rate $g_\text{0,o}$ and the phase shift $\Delta\phi$ experienced by an optical photon due to the number of phonons  $n_\text{ph}$ occupying the mechanical resonator \cite{tsang_cavity_2010}:

\begin{equation} \label{eq_phase_g0}
g_{\text{0,o}}  \sqrt{n_{\text{ph}}} = \frac{\Delta\phi}{\tau}
\end{equation}

where $\tau$ is the lifetime of the photons in the optical cavity for which holds $\tau=1/\kappa_\text{o}$.

The number of phonons that are generated by a microwave signal $P_\mathrm{e}$ at the microwave resonance frequency $\omega_\text{e}$ is given by \cite{jiang2019,Ruedacombs}:

\begin{equation} \label{eq_phon_num}
n_{\text{ph}}= |\mathbf{\Theta}_{3,1}(\omega)|^2 \frac{P_\text{e}}{\hbar \omega_\text{e}},
\end{equation}

where $\mathbf{\Theta}$ is the matrix defined by $[-i\omega \mathbf{I}-\mathbf{A}]^{-1}.\mathbf{B}$ using the definitions in section \ref{seq_Num_Mod} and its element $\mathbf{\Theta}_{3,1}(\omega)$ is given by:

\begin{eqnarray} \label{eq_phon_gen_rate}
\mathbf{\Theta}_{3,1}(\omega) =-\frac{i\,\sqrt{\kappa_\text{e}\eta_\text{e}}\, G_\text{e}\chi_\text{e}\chi_\text{m}}{1+[\chi_\text{m}-\chi_\text{m}^*]\big[G_\text{e}^2(\chi_\text{e}-\chi_\text{e}^*)+G_\text{o}^2(\chi_\text{o}-\chi_\text{o}^*)\big]},
\end{eqnarray}

with the susceptibilities $\chi_j(\omega)$ and $\chi_j^*(\omega)$ defined above.

Finally, the microwave input power $P_\text{e}$ can be related to a sinusoidal peak voltage $V_\text{p}$ by:

\begin{eqnarray} \label{eq_MW_volt}
P_{\text{e}} = \frac{V_\text{p}^2}{2Z_{\text{e}}},
\end{eqnarray}

where $Z_\text{e}$ is the impedance of the microwave input waveguide, in our case $50\,\Omega$.

Combining the equations (\ref{eq_phon_num}) to (\ref{eq_MW_volt}) and solving for the voltage $V_\pi$ that causes a $\Delta\phi$ of value $\pi$ leads to:

\begin{equation} \label{eq_Vpi_mat}
V_\pi(\omega) = \frac{1}{|\mathbf{\Theta}_{3,1}(\omega)|} \frac{\pi \kappa_\text{o}}{g_{\text{0,o}}} \sqrt{\hbar \omega_\text{e} 2 Z_\text{e}}.
\end{equation}

The value for $V_\pi$ is minimal at the mechanical resonance (i.e. $\omega=\omega_\text{m}$) since the conversion of microwave photons to phonons given by $|\mathbf{\Theta}_{3,1}(\omega)|$ is maximum. 

There are two important scenarios we want to consider here where the equations can be significantly simplified. The first scenario is the case of 
perfectly red-detuned pump tones (i.e. $\Delta_\text{e}=\Delta_\text{o}=\omega_\text{m}$) and sideband resolution on the microwave as well as on the optical side (i.e. $\kappa_\text{e}, \kappa_\text{o} \ll \omega_\text{m}$) so that we can use here the simplified expression $\Gamma_j=4G_j^2/\kappa_j=\mathcal{C}_j\gamma_\text{m}$ for the electro- and optomechanical damping rates $\Gamma_j$ with cooperativities $\mathcal{C}_j$. Then equation (\ref{eq_phon_gen_rate}) can be approximated by:

\begin{equation} \label{eq_phon_gen_rate_simp_with_C_o}
\mathbf{\Theta}_{3,1}(\omega_\text{m}) \approx -i 2 \sqrt{\eta_\text{e}} \sqrt{\frac{\Gamma_\text{e}}{(\gamma_\text{m}+\Gamma_\text{e}+\Gamma_\text{o})^2}} = -i 2 \sqrt{\frac{\eta_\text{e}}{\gamma_\text{m}}} \sqrt{\frac{\mathcal{C}_\text{e}}{(1+\mathcal{C}_\text{e}+\mathcal{C}_\text{o})^2}}.
\end{equation}

This expression shows that the number of phonons generated due to a microwave signal is maximum when the conditions $\Gamma_\text{e} = \gamma_\text{m}$ and $\Gamma_\text{o} \ll \gamma_\text{m}$ are fulfilled. This will result then in rate matching between the electromechanical damping rate $\Gamma_\text{e}$ and the effective loss rate of the mechanical resonator $\gamma_\text{m}$ leading to the best phonon conversion \cite{Aspelmeyer2014}. For unmatched rates, a larger part of the microwave photons will be reflected and not converted to phonons. The second condition of having a vanishing $\Gamma_\text{o}$ comes from the fact that the conversion of phonons to optical photons leads to an additional loss channel of the mechanical resonator limiting therefore the achievable value for $n_{\text{ph}}$ and $\Delta\phi$ for a given microwave signal. This shows the difference to the operation mode of an microwave-to-optics converter where the condition $\Gamma_\text{e}=\Gamma_\text{o}\gg 1$ has to be fulfilled for efficient conversion.

Consequently the value for $V_\pi$ can be calculated by:

\begin{equation} \label{eq_Vpi__simp_with_C_o}
V_\pi(\omega_{\text{m}}) \approx \frac{1}{2} \sqrt{\frac{1}{\eta_{\text{e}}}} \sqrt{\frac{(\gamma_{\text{m}}+\Gamma_{\text{m}}+\Gamma_{\text{o}})^2}{\Gamma_{\text{e}}}} \frac{\pi \kappa_{\text{o}}}{g_{\text{0,o}}} \sqrt{\hbar \omega_{\text{e}} 2 Z_{\text{e}}} = \frac{1}{2} \sqrt{\frac{\gamma_{\text{m}}}{\eta_{\text{e}}}} \sqrt{\frac{(1+\mathcal{C}_{\text{e}}+\mathcal{C}_{\text{o}})^2}{\mathcal{C}_{\text{e}}}} \frac{\pi \kappa_{\text{o}}}{g_{\text{0,o}}} \sqrt{\hbar \omega_{\text{e}} 2 Z_{\text{e}}}.
\end{equation}

The second scenario describes our transducer. Here, we do not achieve sideband resolution on the optical side but are in the limit of $\kappa_\text{e} \ll \omega_\text{m}$ and $\kappa_\text{o} \gg \omega_\text{m}$ leading to $\Gamma_\text{e}=\mathcal{C}_\text{e}\gamma_\text{m}$ and $\Gamma_\text{o}\rightarrow 0$. For equation (\ref{eq_phon_gen_rate}) holds then:

\begin{equation} \label{eq_phon_gen_rate_simp}
\mathbf{\Theta}_{3,1}(\omega_\text{m}) \approx -i 2 \sqrt{\eta_\text{e}} \sqrt{\frac{\Gamma_\text{e}}{(\gamma_\text{m}+\Gamma_\text{e})^2}} = -i 2 \sqrt{\frac{\eta_\text{e}}{\gamma_\text{m}}} \sqrt{\frac{\mathcal{C}_\text{e}}{(1+\mathcal{C}_\text{e})^2}}.
\end{equation}

The maximum for this expression is given again for a rate matching of $\Gamma_\text{e}$ and $\gamma_\text{m}$ (neglecting any power dependency of $\eta_\text{e}$ and $\gamma_\text{m}$). Important to note is also that this approximation for $\mathbf{\Theta}_{3,1}$ is valid as long as the optical detuning is smaller than the optical linewidth, i.e. $\Delta_\text{o} \ll \kappa_\text{o}$ and not only at the resonance condition $\Delta_\text{o} = \omega_\text{m}$. 

In this scenario the voltage $V_\pi$ can be calculated by:

\begin{equation} \label{eq_Vpi_simp_Ce}
V_\pi(\omega_\text{m}) \approx \frac{1}{2} \sqrt{\frac{1}{\eta_\text{e}}} \sqrt{\frac{(\gamma_\text{m} + \Gamma_\text{e})^2}{\Gamma_\text{e}}} \frac{\pi \kappa_\text{o}}{g_\text{0,o}}\sqrt{\hbar \omega_\text{e} 2 Z_\text{e}} = \frac{1}{2} \sqrt{\frac{\gamma_\text{m}}{\eta_\text{e}}} \sqrt{\frac{(1 + \mathcal{C}_\text{e})^2}{\mathcal{C}_\text{e}}} \frac{\pi \kappa_\text{o}}{g_\text{0,o}}\sqrt{\hbar \omega_\text{e} 2 Z_\text{e}}.
\end{equation}

Using the set of parameters of our transducer we calculate a minimum value of $V_\pi$ of around $16\,\mu\text{V}$ which we reach for low optical pump power (i.e. $P_\text{o}=92\,\text{pW}$) and a microwave pump power of $P_\text{e}=409\,\text{pW}$. Higher optical pump powers lead to an increase of $V_\pi$ due to the discussed optical heating effects leading to an increasing $\gamma_\text{m}$ and a decreasing $\eta_\text{e}$.  

\subsection{Estimate for the energy-per-bit $E_\text{bit}$}
Another important figure of merit related to the device efficiency is the microwave energy $E_\text{bit}$ required to encode a bit on an optical carrier signal where we assume again the case of phase modulation. As a rough estimate we calculate this value as the ratio of the microwave signal power $P_\pi=V_\pi^2/(2Z_\text{e})$ required to achieve a $\pi$ optical phase shift to the effective bandwidth $\Gamma_\text{conv}$. As discussed above $V_\pi$ is minimal for $\Gamma_\text{e}=\gamma_\text{m}$ which results in a modulation bandwidth of $2\gamma_\text{m}$. When we work at this point for a low power optical signal (i.e. $P_\text{o}=92\,\text{pW}$) our device requires an energy $E_\text{bit}$ of around $1.3\,\text{fJ}$. Note that in this discussion we have neglected the microwave pump power $P_\text{e}$ that is required to achieve the parametric enhancement of the electromechanical coupling rate $G_\text{e}=\sqrt{n_\text{d,e}} g_\text{0,e}$ which equals a value of around $409\,\text{pW}$.

\subsection{Comparison of the values of $V_\pi$ and $E_\text{bit}$ to literature}

\begin{figure*}[t]
\centering
\includegraphics[width=0.7\columnwidth]{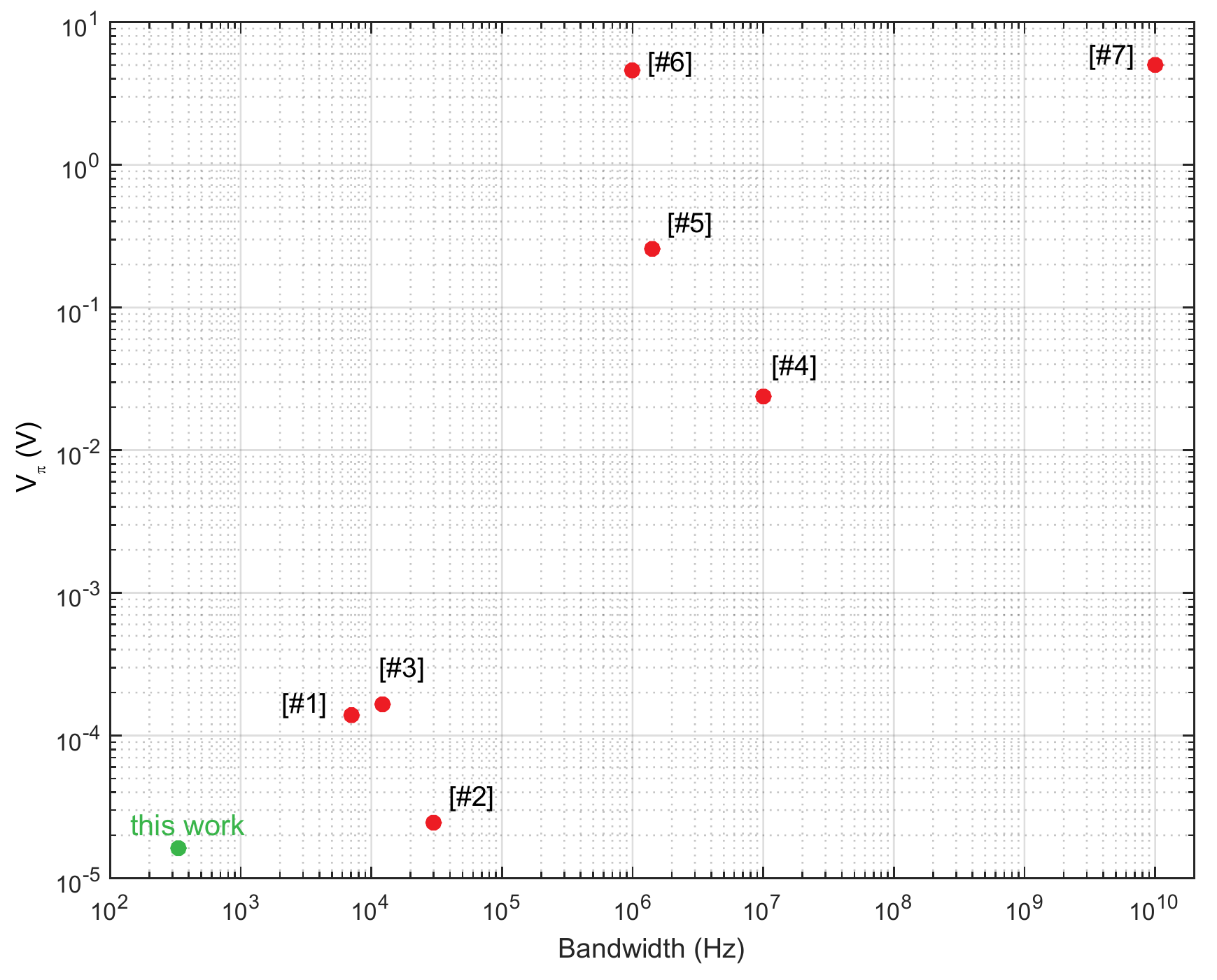}
\caption{\textbf{Performance comparison of phase modulator implementations.} Values of voltage $V_\pi$ and bandwidth of various optical phase modulator platforms. Our implementation offers a new record value of $V_\pi$ which is more than $2 \cdot 10^5$ times better than a commercial Thorlabs electro-optic modulator LN53S-FC (\#7). The price we pay for the very efficient modulation of our triple-resonant system is the low bandwidth. Other reported values are taken from: \#1 Ref.\cite{Bagci2014}, \#2 Ref.\cite{Andrews2014}, \#3 Ref.\cite{Higginbotham2018}, \#4 Ref.\cite{jiang2019}, \#5 Ref.\cite{Ruedacombs}, \#6 Ref.\cite{Shao2019}.
} \label{Fig_Trans_Comp} 
\end{figure*}

Various approaches can be used to implement an optical phase modulator ranging from electro-optomechanics \cite{Bagci2014,Andrews2014,Higginbotham2018} over piezo-optomechanics \cite{Bochmann2013,Balram2016,Shao2019,Ramp2020} to electro-optics \cite{Rueda2016,Fan2018,Ruedacombs}. Our minimum value for $V_\pi$ represents a new record value in the field since it is nearly a factor 9 smaller than the smallest value reported in literature which equals $140\,\mu\text{V}$ \cite{Bagci2014}. Using our analytical equations we are also able to compare our transducer to the most efficient microwave-to-optics converters \cite{Andrews2014,Higginbotham2018} today that are also based on an electro-opto-mechanical system for which no value of $V_\pi$ was reported. For these devices the smallest achieved value equals around $25\,\mu\text{V}$. Different approaches of implementing a converter are much less efficient represented in a much higher value for $V_\pi$. The smallest reported values for a piezo-optomechanical, an electro-optic and a commercial system equal $24\,\text{mV}$ \cite{jiang2019}, $260\,\text{mV}$ \cite{Ruedacombs}, and $5\,\text{V}$ (Thorlabs LN53S-FC), respectively. By minimizing the optical absorption heating leading to a smaller $\gamma_\text{m}$ and improving $\eta_\text{e}$ with an optimized microwave design our device has the potential to achieve an order of magnitude lower $V_\pi$ than reported here. A summary of the voltage $V_\pi$ is shown for various platforms in figure \ref{Fig_Trans_Comp}. Here, $V_\pi$ is plotted against the bandwidth for which the most efficient modulation can be achieved. This is instructive to do since there is usually a dependence between these two parameters, i.e. the most efficient phase modulators use resonances leading to a limited bandwidth.

Our minimum value for the energy-per-bit $E_\text{bit}$ is more than one (nearly two orders) of magnitude more efficient than for a state-of-the-art electro-optic~\cite{Melikyan2014} (piezo-optomechanical~\cite{jiang2019}) modulator. By improving the microwave design leading to larger $\eta_\text{e}$ it will be possible to decrease the value of $E_\text{bit}$ by nearly an order of magnitude leading to values in the sub-femtojoule range.

Important to note in these discussions is that our system requires a parametric amplification on the electro-mechanical side to achieve such low values for $V_\pi$ and $E_\text{bit}$ as mentioned above, i.e. we do need a microwave pump tone. This is not the case for the other approaches. 


\section{Device design}

The transducer device can be divided into the nanomechanical oscillator and the microwave resonator. The nanomechanical oscillator itself consists of the movable string electrodes of the mechanically compliant capacitors of the microwave LC-circuit and the 'zipper'-optomechanical cavity. The design of the string electrodes is based on ref.~\cite{Barzanjeh2019,Barzanjeh2017} whereas the 'zipper'-cavity is inspired by ref.~\cite{Safavi-Naeini2013b}.

The geometry of the device is shown in Fig.~\ref{Fig_Dev_design}a where all important dimensions are highlighted. The nanomechanical oscillator is surrounded by rectangular cutouts etched in the silicon device layer which act as a buckling shield~\cite{Barzanjeh2019} to reduce the effect of membrane buckling due to compressive stress in the silicon. That would lead to out of plane misalignment of our resonator and therefore to a decrease of $g_\text{0,o}/(2\pi)$ and $g_\text{0,e}/(2\pi)$.
\subsection{Optical cavity}
\begin{figure*}[t]
\centering
\includegraphics[width=0.9\columnwidth]{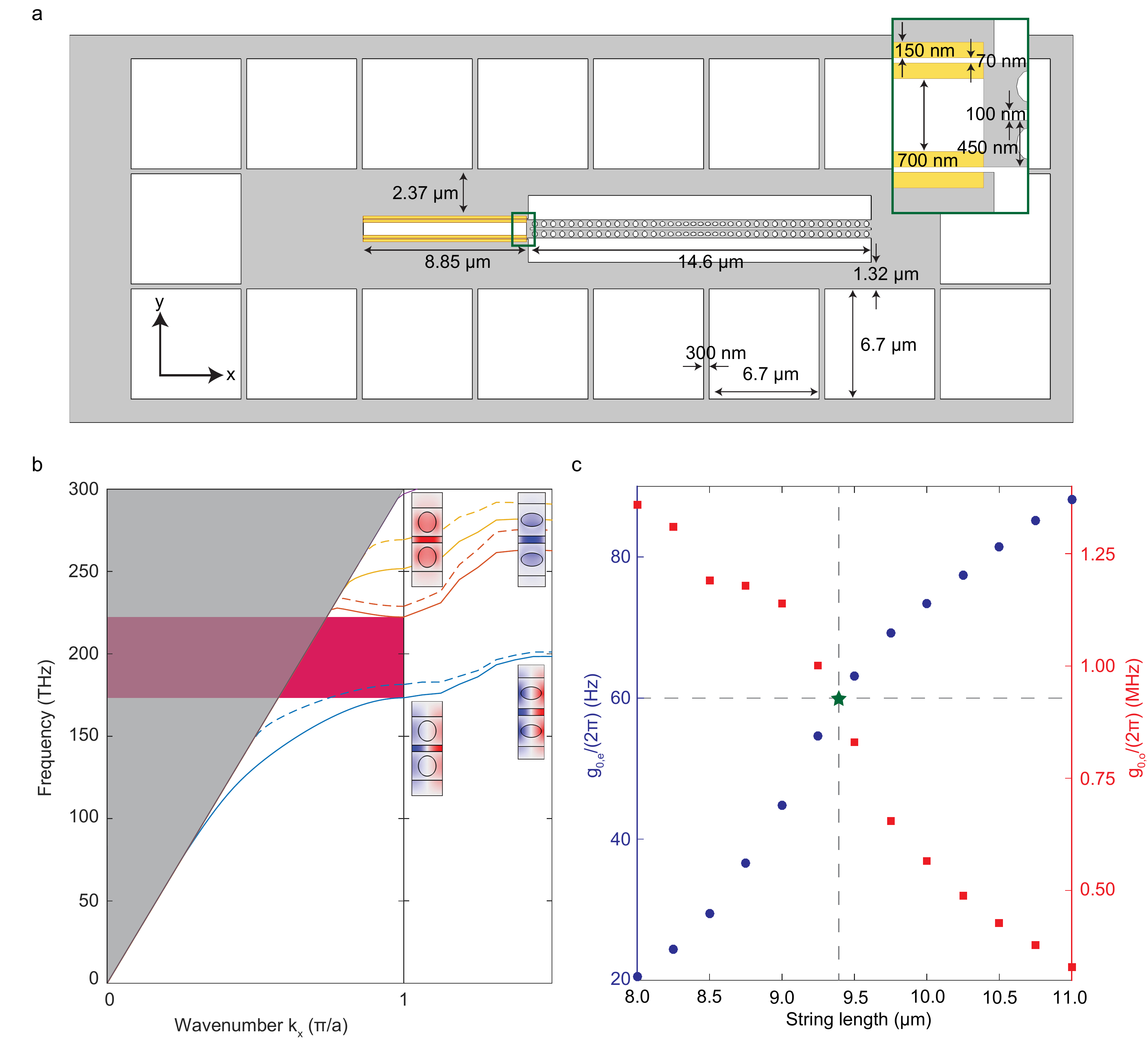}
\caption{\textbf{Device design.} 
\textbf{a}, Geometry of the transducer device. Important dimensions are indicated and the aluminum capacitor electrodes are shown in yellow. Further circuit wiring is not shown. 
\textbf{b}, Photonic band diagram for the optomechanical 'zipper' cavity. The left part of the diagram shows the bands for the mirror region as a function of wavenumber whereas the right part depicts the evolution of the bands towards the cavity defect with a gradual change of hole dimensions. The gray shaded area illustrates the continuum of unguided modes and the pink area highlights the band gap. Solid lines show modes which are symmetric in the y-direction whereas dashed lines show anti-symmetric ones. Insets show the spatial electric field distribution $E_\text{y}(x,y)$ for the first and second y-symmetric mode. 
\textbf{c}, Electromechanical (left y-axis) as well as optomechanical (right y-axis) coupling rate extracted from a series of FEM simulations where the string length of the nanomechanical oscillator was swept between 8 and $11\,\mu\text{m}$. The dashed gray lines indicate the string length chosen for the final design and the corresponding values for $g_{0,\text{e}}/(2\pi)$ and $g_{0,\text{o}}/(2\pi)$.
} \label{Fig_Dev_design} 
\end{figure*}

The optomechanical 'zipper' cavity was designed in two steps. First, the dimensions of the photonic crystal mirror were determined by FEM simulations (COMSOL multiphysics$^\text{\textregistered}$) of the unit cell in such a way that there is large bandgap between the first and the second guided band centered around the desired frequency of $\sim200$\,THz. As a second step, the hole size and the lattice constant were modified to pull up the first mode into the center of the bandgap as shown in Fig.~\ref{Fig_Dev_design}b. This approach leads to a cavity mode with high quality factor. 

The optomechanical 'zipper' cavity is evanescently coupled to a coupling waveguide \cite{Groeblacher2013a}. Consequently, the distance between cavity and the waveguide determines the strength of the optical waveguide coupling rate $\kappa_\text{ex,o}/(2\pi)$. We chose a distance of 400\,nm which should lead to a value of 2.15\,GHz, according to FEM simulations. This value is comparable to prior experimentally observed internal optical cavity loss rates $\kappa_\text{in,o}/(2\pi)$. Unfortunately, we observe a much smaller optical waveguide coupling rate of 0.20\,GHz in the fabricated transducer device. The reason for this deviation is subject of future investigations but first FEM simulations indicate fabrication inaccuracies as the cause.
\subsection{LC circuit}
The microwave resonator is implemented by means of a LC-circuit. The capacitance is realized by two capacitors connected in parallel, each of them consisting of two aluminum electrodes separated by a $\sim70$\,nm gap leading to a capacitance of around 0.43\,fF according to FEM simulations. Each string of the nanomechanical oscillator acts as one electrode of a capacitor and is therefore mechanically compliant. The inductor is implemented by means of a square coil which consists of 48 turns with a pitch of $0.5\,\mu\text{m}$. Its inductance has an analytically calculated value \cite{Mohan1999} of 59.8\,nH. The length of the aluminum wiring between the coil and the capacitors is minimized in order to reduce the stray capacitance of the circuit. Taking into account the simulated values of the two mechanically modulated capacitors with a capacitance of $2 \cdot 0.43 $fF, the coil inductance and the measured microwave resonance frequency yields a stray capacity of $C_\text{s} \approx 3\,\text{fF}$. Our goal was to achieve significant overcoupling for the microwave circuit and we therefore chose the distance of the coil from the microwave feed line to be $9.5\,\mu\text{m}$ which should have led to an extrinsic coupling rate $\kappa_{\text{ex,e}}/(2\pi)$ of $0.9\,\text{MHz}$, a value above the typical intrinsic losses that we experienced in prior experiments for similar structures. However, due to the heating mechanisms described in the main text (Fig.~\ref{Fig_noise}), we were undercoupled when turning on the optical pump (Fig.~\ref{Fig_conversion}e). 

\subsection{Electro- and optomechanical coupling}
Since the properties of the mechanical motion of the nanomechanical oscillator highly depend on its geometry, its design has to be optimized to reach the best electro- and optomechanical coupling rates. As first step, we choose the length of the optical cavity according to two criteria: 1. the number of mirror cells had to be large enough to reach a decent intrinsic quality factor (in this experiment a value of $1.3 \cdot 10^5$ equivalent to an intrinsic cavity loss rate of $\kappa_{\text{in,o}}/(2\pi)=1.42\,\text{GHz}$) and 2. the total length had to be short enough so that we could keep the thermal occupation at the fridge base temperature relatively low. As second step, we adjusted the length of the strings. For this purpose, we conducted a series of FEM simulations for varying string lengths, where we extracted $g_{\text{0,o}}/(2\pi)$ and $g_{\text{0,e}}/(2\pi)$ for the antisymmetric differential in-plane mechanical mode. As described above, the used microwave circuit consists of two parallel mechanically compliant capacitors, each of them with a capacitance of $C_{\text{m}}=0.43\,\text{fF}$.
The electrical frequency shift per displacement $g_{\text{em}}$ is given by the following expression:

\begin{eqnarray}
g_{\text{em}} = -\eta \frac{\omega_\text{e}}{2} \frac{1}{2 C_\text{m}} \frac{\partial C}{\partial u} = -\frac{2 C_\text{m}}{2 C_\text{m} + C_\text{s}} \frac{\omega_\text{e}}{2} \frac{1}{2 C_\text{m}} \frac{\partial C}{\partial u},
\end{eqnarray}%

where $u$ is the mechanical modal amplitude coordinate and $\eta=2 C_{\text{m}}/(2 C_\text{m} + C_\text{s})$ the motional participation ratio. The electromechanical vacuum coupling rate can then be calculated by $g_\text{0,e} = 2 x_{\text{zpf}} g_{\text{em}}$ with $x_{\text{zpf}}=\sqrt{\hbar / 2 m_{\text{eff}} \omega_{\text{m}}}$ as the zero-point amplitude and $m_{\text{eff}}$ as the motional mass. FEM simulations revealed values for $m_{\text{eff}}$ and $x_{\text{zpf}}$ of 1.3\,pg and 24.5\,fm, respectively. The results for the string length sweep are shown in Fig.~\ref{Fig_Dev_design}c. Clearly the electromechanical coupling rate increases with the string length while the optomechanical pendant decreases accordingly. For a string length of $9.4\,\mu\text{m}$ the 'zipper' and the string parts of our nanomechanical resonator have roughly the same effective mass and therefore also their zero-point fluctuations are very similar. As a consequence, both the electro- and the optomechanical coupling rate have a decent value which is important for the transducer since conversion requires similar cooperativity for both processes. For this geometry the mechanical mode has simulated a resonance frequency of 10.88\,MHz and the coupling rates have simulated values of 60\,Hz and 893\,kHz.

\subsection{Mechanical oscillator}

\begin{figure*}[t]
\centering
\includegraphics[width=0.8\columnwidth]{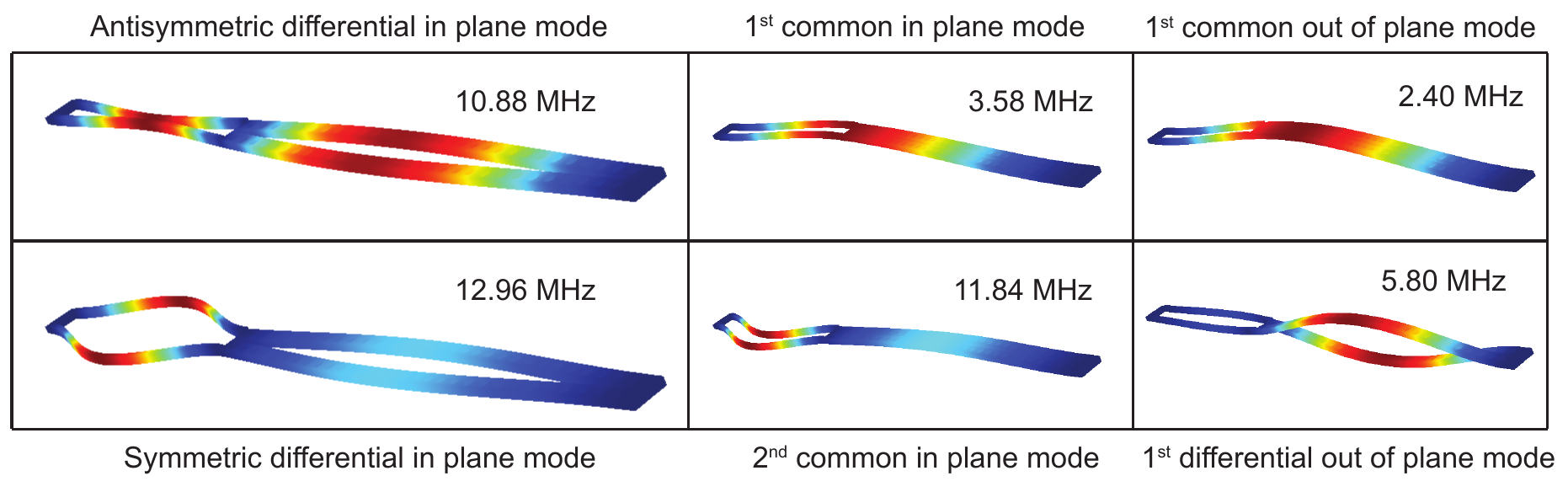}
\caption{\textbf{Simulation of the most relevant mechanical modes.} FEM simulation of the mechanical displacement and the resonance frequency of the first fundamental modes of our nanomechanical oscillator. 
} \label{Fig_Mech_modes} 
\end{figure*}

Since our transducer requires a mechanical motion which modulates maximally the capacitor gap as well as the gap of the optical cavity we included an elastic pinning block which connects the two silicon nanobeams of the 'zipper' at the intersection point with the strings. This pinning block forces the device to feature two in plane differential mechanical modes that create the desired strong electro- and optomechanical interaction and the double clamped tuning fork geometry limits the clamping losses due to elastic wave interference ~\cite{Zhang2015a}. Another design criterion is the frequency gap between the chosen mechanical mode and other existing resonances of the nanomechanical resonator. The goal here is to avoid hybridization of the chosen mechanical mode with other modes which would cause a significant decrease in the electro- and optomechanical coupling rate. For this purpose we simulated the first mechanical resonances of our structure. The result is shown in Fig.~\ref{Fig_Mech_modes}. There are two in plane differential modes which we distinguish by the terms symmetric and antisymmetric depending on if the strings and the beams of the 'zipper' cavity are oscillating in phase or not. Both of these resonances have significant values for $g_{\text{0,o}}/(2\pi)$ and $g_{\text{0,e}}/(2\pi)$ since the capacitor gap as well as the gap between the nanobeams is strongly modulated. 
In the end, we chose to work with the antisymmetric mode since it features slightly bigger coupling rates. Beside these modes there are 3 other family of modes, namely the common in plane, the common out of plane and the differential out of plane modes. However, all of these resonances exhibit negligible electro- and optomechanical coupling. Moreover, all modes are at least 1 MHz away from the mechanical mode we work with. Thus, the other mechanical modes do not influence the performance of the transducer. Also in experiment, we have not seen indications of optomechanical coupling to other modes than to the two differential in plane modes.

\subsection{Device parameter summary}
All important device parameters are summarized in table \ref{Tab_device_param}. For each parameter the design value from simulation as well as the experimentally extracted value is listed, if available.
\begin{table}[ht]
\centering
\begin{tabular}{| c | c | c | c | c |}
\hline
Parameter &  \multicolumn{2}{c}{Simulated value} & \multicolumn{2}{|c|}{Measured value} \\
\hline
\textbf{$\omega_{\text{o}}/(2\pi)$} & 193.874 & $\text{THz}$ & 198.081 & $\text{THz}$  \\
\hline
\textbf{$\Delta_{\text{o}}/(2\pi)$} & - & - & 126 & $\text{MHz}$  \\
\hline
\textbf{$\kappa_{\text{ex,o}}/(2\pi)$} & 2.15 & $\text{GHz}$ & 0.18 & $\text{GHz}$ \\
\hline
\textbf{$\kappa_{\text{in,o}}/(2\pi)$} & 0.02 & $\text{GHz}$ & 1.42 & $\text{GHz}$ \\
\hline
\textbf{$\omega_{\text{e}}/(2\pi)$} &  10.387 & $\text{GHz}$ & \parbox{3.5cm}{10.497 ($P_\text{o}=0\,\text{pW}$) \\ 10.490 ($P_\text{o}=92\,\text{pW}$) \\ 10.478 ($P_\text{o}=1556\,\text{pW}$)} & $\text{MHz}$ \\
\hline
\textbf{$\Delta_{\text{e}}/(2\pi)$} & - & - & 11.84 & $\text{MHz}$  \\
\hline
\textbf{$\kappa_{\text{ex,e}}/(2\pi)$} &  0.9 & $\text{MHz}$ & 1.15 & $\text{MHz}$ \\
\hline
\textbf{$\kappa_{\text{in,e}}/(2\pi)$} & - & - & \parbox{3.5cm}{1.6 ($P_\text{o}=0\,\text{pW}$) \\ 6.1 ($P_\text{o}=92\,\text{pW}$) \\ 13.9 ($P_\text{o}=1556\,\text{pW}$)} & $\text{MHz}$ \\
\hline
\textbf{$L_{\text{coil}}$} & 59.8 & $\text{nH}$ & - & - \\
\hline
\textbf{$C_{\text{s}}$} & 3.0 & $\text{fF}$ & - & - \\
\hline
\textbf{$C_{\text{m}}$} & 0.9 & $\text{fF}$ & - & - \\
\hline
\textbf{$\omega_\text{m}/(2\pi)$} & 10.9 & $\text{MHz}$ & 11.84 & $\text{MHz}$ \\
\hline
\textbf{$\gamma_\text{m}/(2\pi)$} & - & - & \parbox{3.5cm}{15 ($P_\text{o}=0\,\text{pW}$) \\ 164 ($P_\text{o}=92\,\text{pW}$) \\ 355 ($P_\text{o}=1556\,\text{pW}$)} & $\text{Hz}$ \\
\hline
\textbf{$x_{\text{zpf}}$} & 24.5 & $\text{fm}$ & - & - \\
\hline
\textbf{$m_{\text{eff}}$} & 1.3 & $\text{pg}$ & - & - \\
\hline
\textbf{$g_{\text{0,o}}/(2\pi)$} & 893 & $\text{kHz}$ & 662 & $\text{kHz}$ \\
\hline
\textbf{$g_{\text{0,e}}/(2\pi)$} & 60 & $\text{Hz}$ & 67 & $\text{Hz}$ \\
\hline
\end{tabular}
\caption{Table containing a summary of all important device parameters.}
\label{Tab_device_param} 
\end{table}


\section{Experimental setup} \label{app_sec_setup}

\begin{figure*}[t]
\centering
\includegraphics[width=1.0\columnwidth]{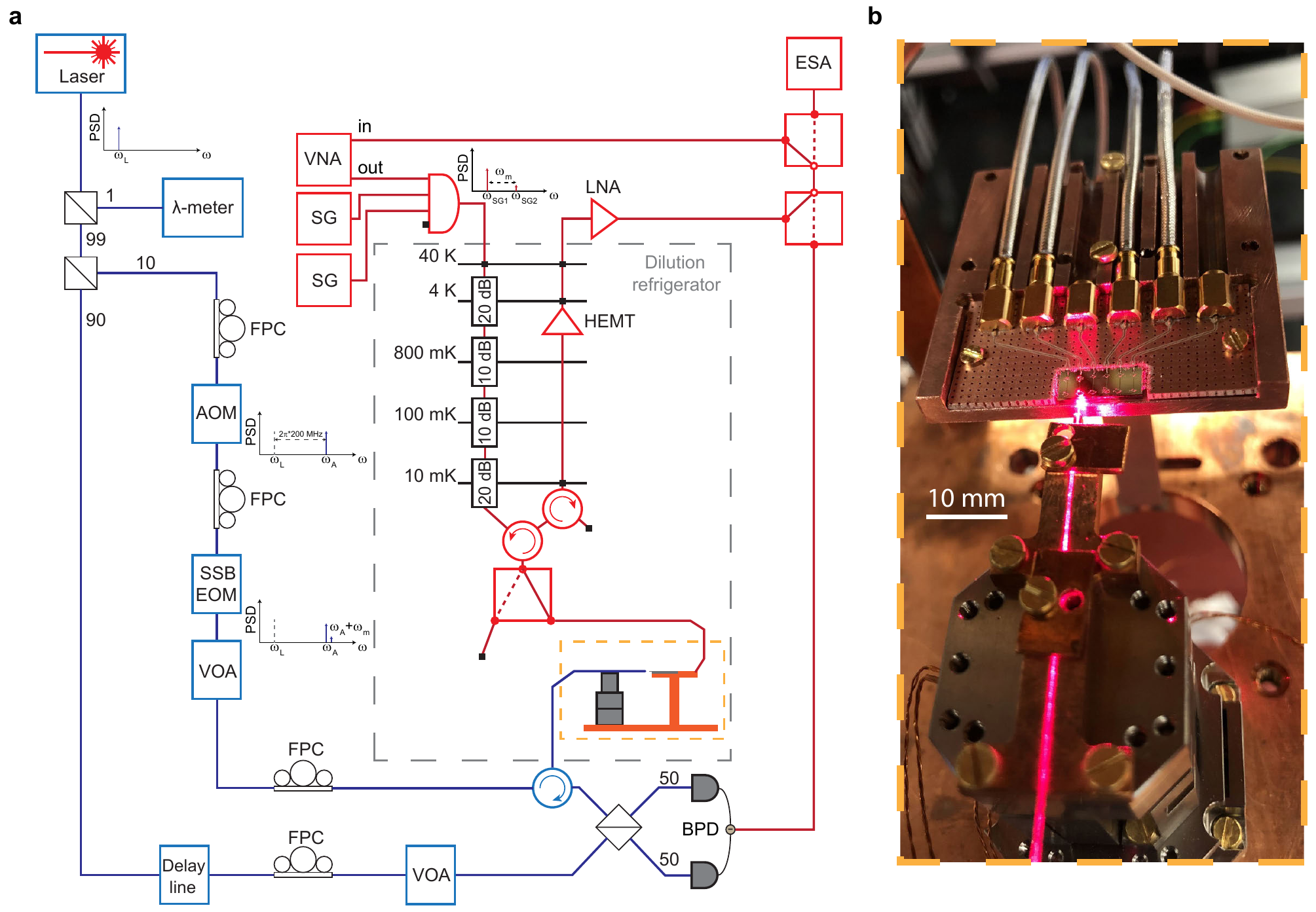}
\caption{\textbf{Optical and microwave conversion measurement setup.} \textbf{a}, Schematic of the optical (blue) and microwave (red) setup used for characterizing the microwave-to-optics converter. In the optical setup the laser light is split into two branches, a high-power local oscillator and a low-power signal arm, for building a high-sensitive heterodyning setup. In the signal arm an AOM is used for shifting the light frequency and a SSB EOM is applied for generating a single weak frequency-shifted probe tone. The light reflected from the sample is recombined with the strong local oscillator and measured on a BPD. In the microwave setup the output of the VNA is combined with the output of two SGs and the signals are sent through a cable in the dilution refrigerator to the sample. The microwave signal reflected from the sample is amplified by a HEMT and a LNA. Two microwave switches allow to choose: first if the reflected microwave or the electrical response from the BPD is analyzed and second if the reflected signal is sent to the VNA or the ESA. The spectral position of all signals are indicated schematically at important positions in the setup. Acronyms: Wavemeter ($\lambda$-meter), Fiber polarization controller (FPC), Acousto-optic modulator (AOM), Single-sideband electro-optic modulator (SSB EOM), Variable optical attenuator (VOA), Balanced photodetector (BPD), Vector network analyzer (VNA), Microwave signal source (SG), High-electron-mobility transistor (HEMT), Low-noise amplifier (LNA), Electronic spectrum analyzer (ESA). \textbf{b}, Photograph showing the alignment of the lensed fiber to the chip using the stack of piezo nanopositioners and the mounting of the chip on the printed circuit board in the dilution refrigerator. 650~nm laser light was sent through the optical measurement system instead of telecom wavelengths to achieve visibility of the optical path to the sample.} \label{Fig_setup}
\end{figure*}

The full measurement setup used for characterizing the microwave-to-optics converter is shown in detail in figure \ref{Fig_setup}a. It consists of two parts, namely an optical (blue color) and a microwave (red color) reflection setup. The converter is mounted on a stage made out of OFC (oxygen-free copper) attached to the mixing chamber plate of a dilution refrigerator (Bluefors LD250) which is kept at a temperature of $\sim$50\,mK, if not specified differently, e.g. in section \ref{seq_MW_Cal} and \ref{seq_opt_Cal}.

As light source for our optical setup we use a fiber-coupled tunable external-cavity diode laser (Santec TSL-550 type A) operated around a frequency of $\omega_{\text{o}}/(2\pi)=198.0815\,\text{THz}$. Using a 99:1 fiber coupler a small fraction of the light is sent to a wavemeter ($\lambda$-meter, Newport WM-1210) for frequency stabilization. The remaining light is divided by a 90:10 fiber coupler into two branches, a low-power signal and a high-power local oscillator arm, required for building an optical heterodyning setup. In the signal arm an acousto-optic modulator (Gooch \& Housego T-M200-0.1C2J-3-F2P) is used to shift the light frequency by 200\,MHz. Afterwards the light is sent through an single-sideband electro-optic modulator (SSB EOM, Thorlabs LN86S-FC) to generate a single small (approximately 20\,dB smaller) optical probe tone detuned by the mechanical frequency $\omega_\text{m}/(2\pi)$. Note that the SSB EOM is operated in such a way that the carrier is not suppressed. Subsequently, the optical signal passes through a variable optical attenuator (VOA, HP8156A) to control the light level that is sent to the sample. Finally, the light is sent to a circulator which routes the light into the dilution refrigerator where a lensed fiber mounted on a stack of attocube$^\text{\textregistered}$ piezo nanopositioners is used for end-fire coupling to the desired device on the mounted chip with a single-pass coupling efficiency of 64\,\% (see Fig.\,\ref{Fig_setup}b). The light reflected by the sample is recombined on a 50:50 fiber coupler with the local oscillator signal whose amplitude is kept at roughly $800\,\mu \mathrm{W}$ with an additional variable optical attenuator. Important to note is that the length of both arms were matched to achieve the lowest noise level. The recombined signal is measured eventually on a balanced photodetector (BPD, Thorlabs PDB470C-AC).

In the microwave setup the signals of three devices, i.e. the output port of a vector network analyzer (VNA, Rohde \& Schwarz ZNB 20) and two microwave signal sources (SG, Rohde \& Schwarz SGS 100A and Rohde \& Schwarz SMA 100B), are first combined by a power combiner and then sent together through a cable to the sample in the dilution refrigerator. The signal is attenuated at every temperature stage to eliminate Johnson-Nyquist noise. Using a circulator the microwave signal is routed to the sample which is mounted on a printed circuit featuring coplanar microwave waveguides to direct the RF signal to the chip (see Fig.~\ref{Fig_setup}b), or alternatively to a low-temperature 50 Ohm termination by employing a microwave switch mounted also at the mixing chamber plate. On the output side a second circulator is used to isolate the sample from thermal noise coming from the hotter stages above. After this isolator the reflected microwave signal is sent to two amplifiers, i.e. a high-electron-mobility transistor (HEMT, Low-noise factory LNC6-20C) mounted at the 4K-stage in the refrigerator and a low-noise amplifier (LNA, Agile AMT-A0067) positioned outside of the cryostat.

Two microwave switches allow us to decide which signal we want to analyze. The first switch grants us the possibility to choose between the reflected microwave signal or the electronic response of the balanced photodetector. The second switch routes this signal then either to the input port of the VNA or to an electronic spectrum analyzer (ESA,  Rohde \& Schwarz FSW 26). In conclusion, this allows us to make three types of measurements: 1. measure the microwave resonator with the VNA, 2. spectrally analyze the reflected microwave signal or 3. spectrally analyze the reflected optical signal.

\section{Characterization}
\subsection{Resonator measurements}

\begin{figure*}[t]
\centering
\includegraphics[width=0.9\columnwidth]{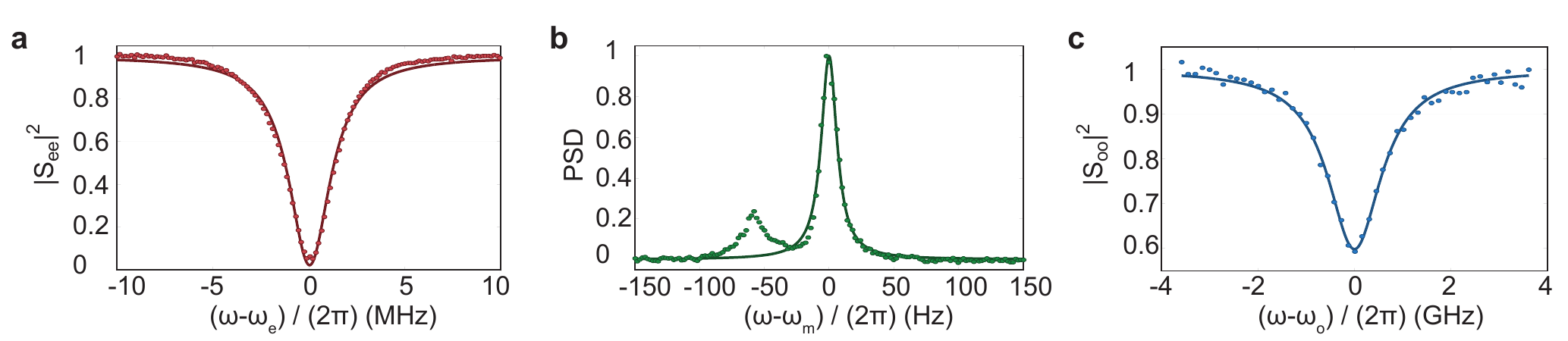}
\caption{\textbf{Cavity and resonator measurements.} \textbf{a}, Reflection around the microwave resonance frequency of $\omega_\text{e}/(2\pi)=10.5\,\text{GHz}$. \textbf{b}, Power spectral density of the reflected microwave signal peaking at $\omega_\text{m}/(2\pi)=11.84\,\text{MHz}$ and \textbf{c}, Optical reflection around the cavity resonance of $\omega_\text{o}/(2\pi)=198.081\,\text{GHz}$. Dots represent the experimental data whereas the lines show the Lorentzian fits. 
} \label{Fig_CavRes_reflections} 
\end{figure*}

We characterize the microwave and the optical cavity at a base temperature of 50\,mK in our dilution refrigerator by sweeping a weak probe tone in the corresponding frequency range over the used resonances and extract the scattering parameters $|S_\text{ee}|^2$ (microwave reflection) and $|S_\text{oo}|^2$ (optical reflection) of the converter. These reflection measurements show a Lorentzian dip around the resonances. The microwave reflection (Fig.~\ref{Fig_CavRes_reflections}a) has a resonance (with optical pump off) around $\omega_\text{e}/(2\pi)=10.5\,\text{GHz}$ with a total loss rate of $\kappa_\text{e}/(2\pi)=2.7\,\text{MHz}$ and a waveguide coupling rate of $\kappa_\text{ex,e}/(2\pi)=1.15\,\text{MHz}$ leading to a coupling ratio of $\eta_\text{e}=\kappa_\text{ex,e}/\kappa_\text{e}=0.43$. In contrast, the optical resonance (Fig.~\ref{Fig_CavRes_reflections}c) at $\omega_\text{o}/(2\pi)=198.081\,\mathrm{THz}$ is much shallower due to the much smaller coupling ratio of $\eta_\text{o}=0.11$ connected with a $\kappa_\text{o}/(2\pi)=1.60\,\text{GHz}$ and a $\kappa_\text{ex,o}/(2\pi)=0.18\,\text{GHz}$. The power spectral density of the reflected photons reveals the mechanical resonance frequency at $\omega_\text{m}/(2\pi)=11.8424\,\text{MHz}$ and a mechanical decay rate of $\gamma_\text{m}/(2\pi)\approx15\,\text{Hz}$ (Fig.~\ref{Fig_CavRes_reflections}b, optical pump off).

\subsection{Calibration}

As described in ref.~\cite{Andrews2014}, there is a way to extract the conversion efficiency without the need to know explicitly the gain and attenuation in the measurement setup. In detail, the product of the resonant microwave-to-optics ($P_\text{oe}(\delta_\text{e}=0)$) and optics-to-microwave transduced ($P_\text{eo}(\delta_\text{o}=0)$) power is divided by the product of the off-resonantly measured microwave ($P_\text{ee}(|\delta_\text{e}|\gg\kappa_\text{e})$) and optical reflection power ($P_\text{oo}(|\delta_\text{o}|\gg\kappa_\text{o})$). The square root of this value yields the desired mean bidirectional photon number conversion efficiency, i.e. $|S_\text{eo,oe}|^2 = P_\text{eo} P_\text{oe} / (P_\text{ee} P_\text{oo})$.

However, quantifying the noise quanta added during the conversion requires knowledge of the gain and the attenuation in the microwave as well as the optical measurement. In the following two subsections the calibration procedure for extracting these parameters is explained. 








\subsubsection{Microwave measurement system} \label{seq_MW_Cal}

Calibrating the microwave measurement system refers to extracting the most important sample parameter on the microwave side, namely the electromechanical coupling rate $g_\text{0,e}$, the effective microwave gain $\mathcal{G}_{\text{setup,e}}$ and the noise added by the chain of amplifiers used to measure the reflected microwave signal $n_{\text{add,setup,e}}$ (see section \ref{app_sec_setup} for details about the setup). This procedure involves a multi-step process consisting of quantifying first $g_{\text{0,e}}$, calculating then $n_{\text{add,setup,e}}$ and the attenuation on the microwave input side and finally deriving the gain of the chain of amplifiers on the output side.

To extract $g_{\text{0,e}}$ we measured the thermal noise power spectrum of the microwave reflection as a function of fridge temperature $T_{\text{fridge}}$. Specifically, we swept $T_{\text{fridge}}$ from the base temperature of around 7\,mK to around 407\,mK while recording the noise power spectrum for a weak microwave pump tone so that the electromechanical backaction can be neglected. Each noise measurement was once done for red- and blue-detuning, i.e. $\Delta_\text{e}=\pm\omega_\text{m}$, to prove that we are in the low cooperativity limit and to generate two complementary sets of data.

It can be shown that in this limit the noise power spectrum $S_\text{e}(\omega)$ is described by the following equation (see ref.~\cite{Fink2016} for a detailed derivation):

\begin{eqnarray}\label{eq_MW_noise}
\frac{S_\text{e}(\omega)}{P_\text{r}} = \mathcal{O}_{\text{e}} + \frac{64 n_\text{m} \kappa_{\text{ex,e}}^2 \gamma_\text{m} g_\text{0,e}^2}{(4 \Delta_{\text{e}}^2 + (\kappa_{\text{e}} - 2 \kappa_{\text{ex,e}})^2)(\kappa_{\text{e}}^2 + 4 (\Delta_{\text{e}}-\omega)^2)(\gamma_\text{m}^2 + 4 (\omega_\text{m} - \omega)^2)}.
\end{eqnarray}

By normalizing the spectrum $S_\text{e}(\omega)$ by the reflected microwave pump power $P_\text{r}$ the dependence on $\mathcal{G}_{\text{setup,e}}$ of the measurement setup drops out. If we are now able to quantify the mechanical occupation $n_\text{m}$, equation (\ref{eq_MW_noise}) will allow us to extract $g_{\text{0,e}}$ since all other parameters are known from separate measurements.

We have knowledge about $n_\text{m}$, when we know the effective temperature $T_\text{m}$ of the mechanical resonator, since they are related to each other by the Bose-Einstein statistics. In the easiest scenario we can set $T_\text{m}$ equal to $T_{\text{fridge}}$ which will be the case if the sample has thermalized with the mixing chamber plate. According to experience this will not be the case for temperatures close to the base temperature of 7\,mK but only for elevated temperatures. To verify that the sample is thermalized with the refrigerator we extract $g_{\text{0,e}}$ for a range of mixing chamber temperatures. When the extracted value for the electromechanical coupling is constant with temperature while assuming that $T_\text{m}=T_{\text{fridge}}$, the sample is thermalized with its environment and the value of $g_{\text{0,e}}$ is trustworthy.

\begin{figure*}[t]
\centering
\includegraphics[width=0.6\columnwidth]{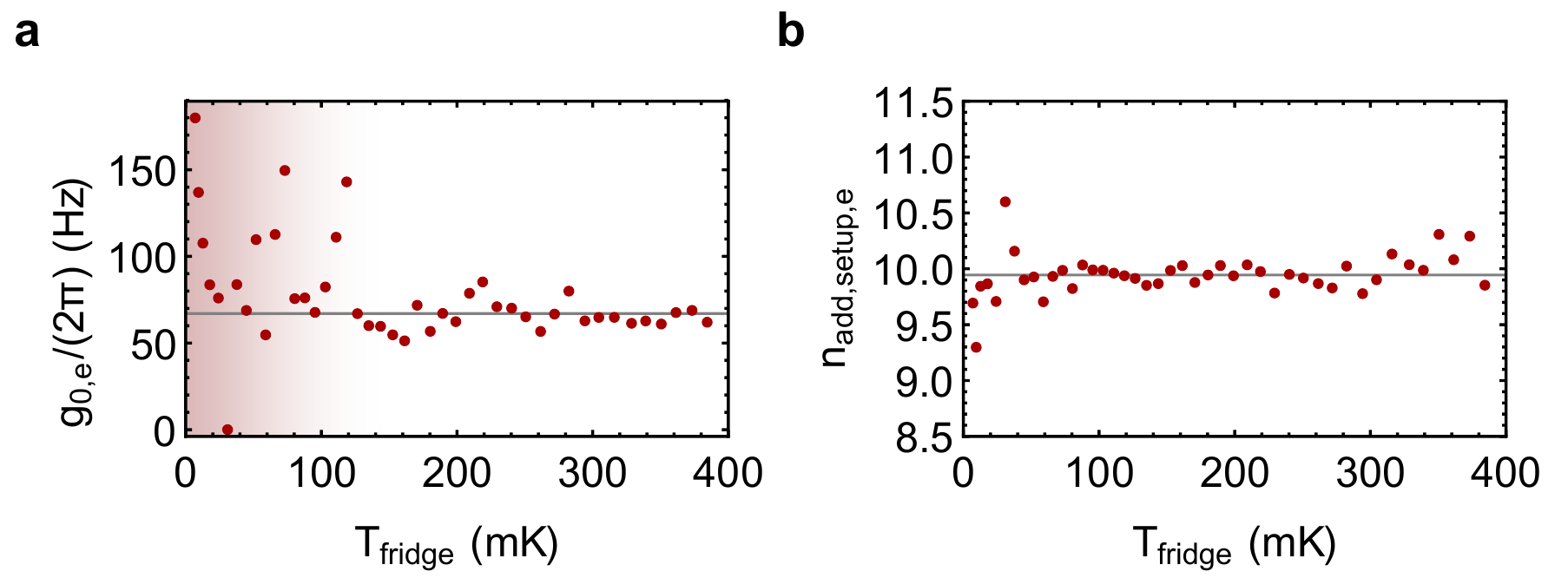}
\caption{\textbf{Microwave calibration.} \textbf{a}, Extracted $g_{\text{0,e}}/(2\pi)$ for a series of thermal noise spectra with a red-detuned microwave pump tone while sweeping $T_{\text{fridge}}$ and inherently assuming that the sample is thermalized to its environment (low pump power limit). The horizontal gray line depicts the average value. At low temperatures the sample is not thermalized to the fridge temperature (red shaded area). \textbf{b}, Added noise from the measurement chain to a detected microwave signal extracted from Eq.~\ref{eq_MW_noise_back1}. Also here the horizontal gray line shows the average value.}\label{Fig_MW_cal}
\end{figure*}

Fig.~\ref{Fig_MW_cal}a shows the extracted values of $g_{\text{0,e}/(2\pi)}$ for varying fridge temperatures. The value converges for temperatures above 150\,mK to a mean value of around $67\,\text{Hz}$. For low temperatures the value varies strongly which is because the assumption $T_\text{m}=T_{\text{fridge}}$ is not valid anymore. A huge variance in the noise response for low temperatures and in the low cooperativity limit is a well known problem~\cite{Wollman2015} but - as we also observe in experiment - the value stabilizes for stronger pumping or higher temperatures. Here, we only consider the electromechanical coupling rate extracted from the red-detuned measurements because for the blue-detuning $g_{\text{0,e}}$ was showing consistently a higher value which we attribute to a small amount of parametric gain already having an effect at low cooperativities. This measurement indicates that the sample is only thermalized to the mixing chamber plate for temperatures above 150\,mK.

As next step we extract the noise $n_{\text{add,setup,e}}$ added by the amplifier chain in the microwave output line. For this purpose we examine the background level $\mathcal{O}_{\text{e}}$ of the thermal noise spectra which is described by the following relation (see again ref.~\cite{Fink2016} for a detailed derivation):

\begin{eqnarray}\label{eq_MW_noise_back1}
\mathcal{O}_{\text{e}} = (1+n_{\text{add,setup,e}})\frac{4\kappa_{\text{ex,e}}}{n_{\text{d,e}}(4\Delta_{\text{e}}^2+(\kappa_{\text{e}}- 2 \kappa_{\text{ex,e}})^2)}.
\end{eqnarray}

Important to note is that we need to know the microwave intracavity photon number $n_{\text{d,e}}$ to be able to extract $n_{\text{add,e}}$ which in turn requires knowledge of the exact attenuation $\mathcal{A}_{\text{e}}$ of the microwave input line in the dilution refrigerator. To gain knowledge of this parameter, we performed independent electromagnetically induced transparency (EIT) spectroscopy measurements which allowed us to extract the term $\sqrt{n_{\text{d,e}}} g_{\text{0,e}}$ (measurements are not shown here). The knowledge of $n_{\text{d,e}}$ as function of microwave pump power provided us with the possibility to calculate $\mathcal{A}_{\text{e}}$, which equals roughly 76.8\,dB in the considered frequency range. This value together with equation (\ref{eq_MW_noise_back1}) allowed us to back out the added microwave noise $n_{\text{add,e}}$ shown in Fig.~\ref{Fig_MW_cal}b. The average value for temperatures above 150\,mK equals 9.9.

Alternatively, the background $\mathcal{O}_{\text{e}}$ can also be described by the following equation (see ref.~\cite{Barzanjeh2017} for a detailed derivation):

\begin{eqnarray}\label{eq_MW_noise_back2}
\mathcal{O}_{\text{e}}(\omega) = \hbar \omega 10^{\mathcal{G}_{\text{setup,e}}/10} (1+n_{\text{add,setup,e}})
\end{eqnarray}

where $\mathcal{G}_{\text{setup,e}}$ is the effective gain of the output side of the microwave setup in dB.

Since the only unknown in this equation is $\mathcal{G}_{\text{setup,e}}$ we solve it for this parameter which has a value of around 64.1\,dB.

Thus, we quantified the electromechanical coupling rate and we now know all important parameters of our microwave setup by using this self-consistent calibration method.

\subsubsection{Optical measurement system} \label{seq_opt_Cal}

The procedure for calibrating the optical measurement system, i.e. determining the effective optical gain $\mathcal{G}_ {\text{setup,o}}$, the added noise $n_{\text{add,setup,o}}$ and from these values the optomechanical coupling rate $g_{\text{0,o}}$, is similar from the approach used for the microwave side described above. In essence, we perform a series of thermal noise measurements for varying refrigerator temperature $T_{\text{fridge}}$ and fit these measurements with a numerical model. Eq.~\ref{eq_MW_noise} cannot be applied because it assumes sideband resolution.

Since our optical detection system is not positioned inside the dilution refrigerator, we are able to measure directly the gain $\mathcal{G}_{\text{setup,o}}$ of the optical heterodyning setup. This gain describes the amplification in the process of converting the optical signal coming from the sample to the electrical signal measured on the electrical spectrum analyzer. For the purpose of quantifying $\mathcal{G}_{\text{setup,o}}$ we send a well defined amount of optical power in the signal branch and interfere it with a local oscillator power of $\mathrm{800\,\mu W}$ which is the same power as used in all conversion measurements. The combined signal is measured with the balanced photodetector and finally spectrally analyzed by the electronic spectrum analyzer. By dividing the amplitude of the 200\,MHz peak (shift frequency of the acousto-optic modulator) of the electric power spectral density by the optical power reflected from the sample we calculate the gain $\mathcal{G}_{\text{setup,o}}$ which has a value of around 17.9\,dB.

\begin{figure*}[t]
\centering
\includegraphics[width=0.65\columnwidth]{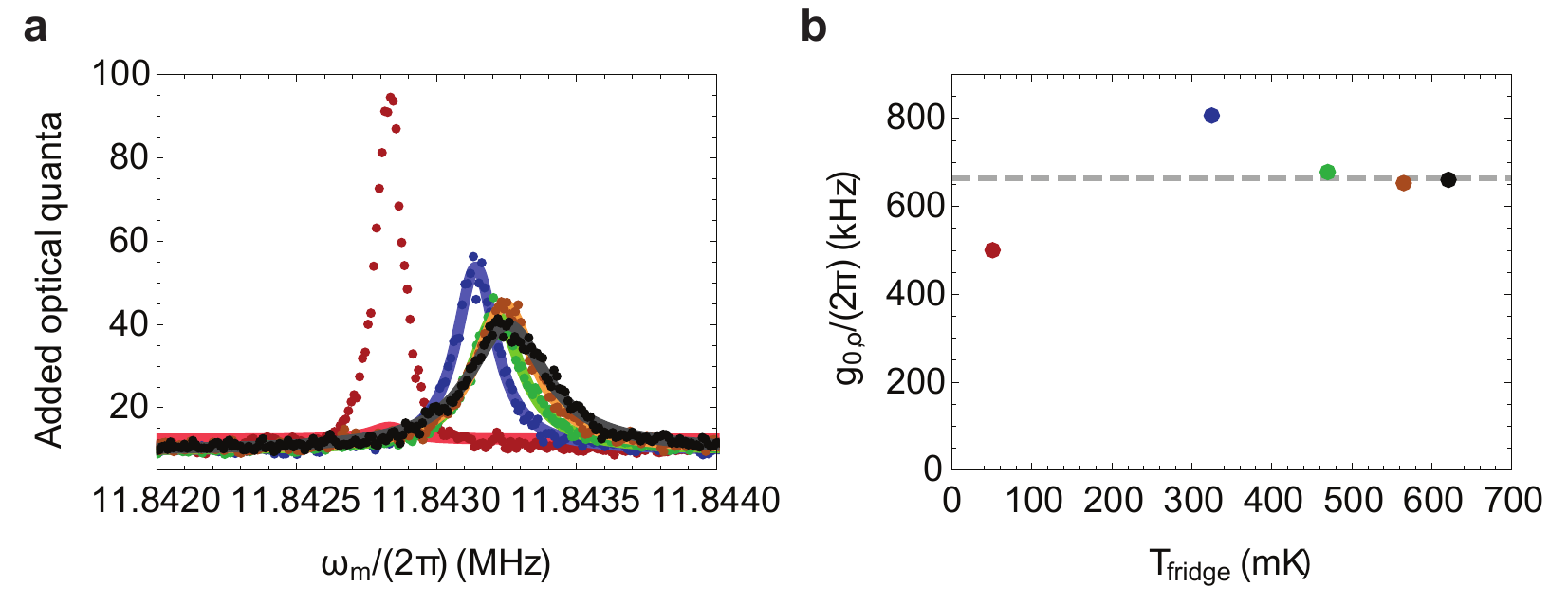}
\caption{\textbf{Extraction of the optomechanical coupling rate $g_{\text{0,o}}$.} \textbf{a}, Measured (dots) and modeled (lines) thermal noise spectra of the optical reflection in unit of number of photons. The temperature $T_{\text{fridge}}$ is swept from 51 to 621\,mK in 5 steps: 51\,mK (red color), 325\,mK (blue color), 469\,mK (green color), 565\,mK (orange color) and 621\,mK (black color). \textbf{b}, Values of the optomechanical coupling rate $g_{\text{0,o}}$ extracted from this measurement series assuming the sample is thermalized to the mixing chamber plate. The horizontal gray dashed line depicts the average value for the three highest temperatures.
} \label{Fig_Opt_cal} 
\end{figure*}

As next step we measure the thermal noise power spectrum on the optics side as a function of refrigerator temperature $T_{\text{fridge}}$. Specifically, we sweep $T_{\text{fridge}}$ from a temperature of around 51\,mK to around 621\,mK while recording the noise power spectrum for a weak optical pump tone, so that the optomechanical backaction can be neglected. The measured electrical power spectral density is converted to the units of number of photons emitted from the optical cavity by dividing by the effective gain $\mathcal{G}_{\text{setup,o}}$ and by the optical single photon energy. The corresponding measured power spectral density are shown in Fig.~\ref{Fig_Opt_cal}a. Three different trends with increasing temperature can be observed in the experimental data. First, the amplitude of the spectrum decreases by more than a factor 2 by going from 51 to 325\,mK from whereon it stays rather constant. This maybe unexpected behavior is connected with the second trend, namely that the spectra broaden with increasing $T_{\text{fridge}}$, i.e. the value of $\gamma_\text{m}$ gets larger as reported in the main text. As a consequence, the optomechanical cooperativity decreases which results additionally in a decrease of the amplitude of the spectra. Third, the center frequency of the power spectral density blueshifts with increasing temperature. 

The expected number of thermal noise photons emitted by our optical cavity can be described by our numerical model where the electromechanical interaction has been switched off (see section \ref{seq_Num_Mod} for details). Assuming now that the sample is thermalized to the mixing chamber plate, we are able to fit the numerical model to the experimental data by using only $\mathrm{g_{0,o}}$, which mostly influences the amplitude of the peak, and an offset $\mathcal{O}_{\text{o}}$ as fit parameters while all other parameter values are fixed by independent measurements. The results of these fits are shown in Fig.~\ref{Fig_Opt_cal}a together with the experimental data. It can be seen that the model is able to quantitatively represent all measured thermal noise spectra except for the lowest temperature where the fit obviously fails. This failure is related to the fact that the sample is not thermalized to the mixing chamber plate for these low temperatures and the assumption $T_{\text{fridge}}=T_\text{m}$ breaks down as discussed in section \ref{seq_MW_Cal}. In essence, the model is not able to represent the large amplitude of the noise spectra with a reasonable value of $g_{\text{0,o}}$ since the effective temperature $T_\text{m}$ of the mechanics is much larger than assumed. The fitted values for $g_{\text{0,o}}/(2\pi)$ as function of $T_{\text{fridge}}$ are shown in Fig.~\ref{Fig_Opt_cal}b. For the two lowest temperatures the fit values do not represent the actual optomechanical coupling rate of our device due to thermalization issues and therefore $T_{\text{fridge}} \neq T_\text{m}$. In contrast, the fitted values for higher fridge temperatures are similar. Thus, we take the mean value of these three fits which equals 662\,kHz as the actual value for $g_{\text{0,opt}}/(2\pi)$. Please note that the minimum temperature at which the sample is thermalized to the fridge is different to Fig.~\ref{Fig_MW_cal}, because the optical pump leads to absorption heating, as discussed in the main text.

Since the fitted background $\mathcal{O}_{\text{o}}$ is already in the units of number of photons we can directly relate it to the number of added noise photons $n_{\text{add,setup,o}}$ caused by our imperfect detection:

\begin{eqnarray}\label{eq_opt_noise_back}
\mathcal{O}_{\text{o}}=1+n_{\text{add,setup,o}}.
\end{eqnarray}

Thus, we can calculate the number of added noise photons by subtracting the vacuum noise of one from the fitted values of $\mathcal{O}_{\text{setup,o}}$. This results in an average value of 8.8 for $n_{\text{add,setup,o}}$.

Alternatively to the approach above, the number of added noise photons can be extracted by modeling our measurement setup as a beam splitter obeying the following input-output formalism:

\begin{eqnarray}\label{eq_in_out1}
a(\omega) = \sqrt{\eta_{\text{qe}}}s(\omega) + \sqrt{1-\eta_{\text{qe}}}v(\omega) 
\end{eqnarray}

where $a(\omega)$ and $s(\omega)$ are the annihilation operators for the optical field at the balanced photodetector and the sample, respectively, $v(\omega)$ is the annihilation operator for an added thermal noise state and $\eta_\text{qe}$ represents the quantum efficiency of our measurement. In essence, this equation describes that in the case of a perfect measurement, i.e. $\eta_\text{qe}= 1.0$, there is no noise added to our output signal, but in all other case there is.

From this equation we can now derive the single sided power spectral density $A_\text{noise}(\omega)$ as measured by the electronic spectrum analyzer:

\begin{eqnarray}\label{eq_PSD}
A_\text{noise}(\omega) = \hbar\omega \int_{-\infty}^{\infty} \left\langle a^{\dagger}(\omega') a(\omega') \right\rangle d\omega'.
\end{eqnarray}

Using this definition equation (\ref{eq_in_out1}) can be transformed into the following form:

\begin{eqnarray}\label{eq_in_out2}
A_\text{noise}(\omega) = \eta_\text{qe} S_\text{noise}(\omega) + (1-\eta_\text{qe}) V_\text{noise}(\omega).
\end{eqnarray}

Here, the quantity $S_\text{noise}(\omega)$ equals the spectrum emitted directly from the optical cavity, whereas the last term $V_\text{noise}(\omega)$ describes the power spectral density of the noise that we add to our measurement.

By dividing equation (\ref{eq_in_out2}) by $\eta_\text{qe}$ we can directly see the relation between the quantum efficiency of the measurement and the added noise photons: 

\begin{eqnarray}\label{eq_in_out3}
\frac{A_\text{noise}(\omega)}{\eta_\text{qe}} = S_\text{noise}(\omega) + \frac{1-\eta_\text{qe}}{\eta_\text{qe}}V_\text{noise}(\omega).
\end{eqnarray}

In this equation the quantities $A_\text{noise}(\omega)/\eta_\text{qe}$ and $S_\text{noise}(\omega) $ represent noise spectra with the background observed in the experiment and a background of one photon coming from vacuum noise, respectively. Consequently, the last term equals directly the added noise photon number $n_{\text{add,setup,o}}$.

To extract $n_{\text{add,setup,o}}$ we first have to find the quantum efficiency which we achieve by fitting our numerical model to the experimental data that is normalized to its noise floor. This results in a value of 0.102 for $\eta_\text{qe}$. Inserting this number into the last term of equation (\ref{eq_in_out3}) and assuming a thermal state with an average photon number of one yields an added noise photon number $n_{\text{add,o}}$ of 8.8 which fits to the observed background level. Thus, our detection system can be interpreted as an ideal measurement setup with a quantum efficiency of 0.102, where a thermal state $\left|1\right\rangle$ is added as noise leading to 8.8 added noise photons. This interpretation is consistent with the calibration method mentioned before using the measured effective gain of our heterodyning setup (Fig.~\ref{Fig_Opt_cal}).

\subsection{Noise and heating} 

\subsubsection{Microwave pump} 

Analogue to Fig.~\ref{Fig_noise}d in the main text, the microwave and optical output noise of the transducer can be plotted with respect to electrical pump power (see Fig.~\ref{Fig_ResNoise}a). For a fixed optical pump power ($P_\text{o}=625\,\text{pW}$), the output noise of the microwave resonator $n_\text{add,e}$ increases with electrical pump power because of an increasing photon-phonon coupling rate $\Gamma_\text{e}$. In turn, the optical output noise $n_\text{add,o}$ decreases with increasing electrical pump powers because of an increasing $\Gamma_\text{e}$, until microwave and optical output noise intersect for matching cooperativities, as can also be seen in the main text (Fig.~\ref{Fig_noise}d). As shortly discussed there, this proves that optical and mechanical output share the same bath. As long as the mechanical noise is the main contribution, the output noise is mainly determined by Eq. \ref{coeff}d. The absolute square of coefficient $\alpha_{j,\text{m}}$ describes the relation between mechanical occupation and photon noise and is proportional to the individual cooperativities. Thus, if on the one hand $C_\text{e} \approx C_\text{o}$ respectively $4G^2_\text{e}/\kappa_\text{e} \approx 4G^2_\text{o}/\kappa_\text{o}$, and on the other hand $n_\text{add,e} \approx n_\text{add,o}$, the mechanical occupation $n_\text{m}$ must be equal for optomechanical and electromechanical resonator.

\subsubsection{Microwave resonator} 

\begin{figure*}[t]
\centering
\includegraphics[width=0.6\columnwidth]{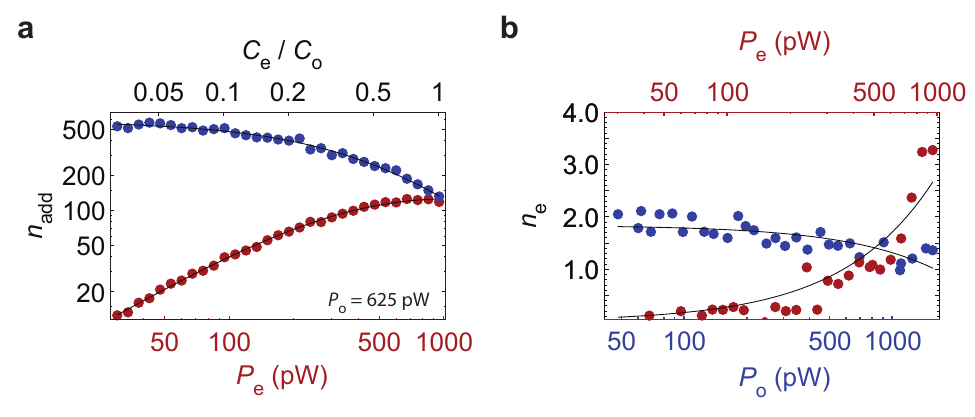}
\caption{\textbf{Transducer output noise and microwave resonator noise} \textbf{a}, Microwave (red, $n_\text{add,e}$) and optical (blue, $n_\text{add,o}$) added noise photons at the output with respect to microwave pump power and a constant optical pump power of 625\,pW (output noise from $n_\text{m}$ fits to our model as black curves) \textbf{b}, Resonator broadband noise with respect to both pumps. The microwave resonator noise $n_\text{e}$ increases linearly with microwave pump power ($P_\text{o}=$~625\,pW, black line is a fit to $2.8 \cdot 10^{-3}P_\text{e}$). In contrast, we attribute the slight decrease with increasing optical pump power ($P_\text{e}=$~601\,pW, fit to $1.8 -5.2\cdot 10^{-4}P_\text{o}$ as black line) to the effect of optical heating on the microwave resonator (see Fig.~\ref{Fig_conversion}e in the main text).
} \label{Fig_ResNoise} 
\end{figure*}

Apart from the optics-related boost of mechanical noise and thereby noise photons added in the frequency-range of the transducer bandwidth, we could also observe a minor contribution from broadband resonator noise $n_\text{e}$ (see Fig.~\ref{Fig_ResNoise}b). It linearly increases with electrical power. The small decrease we observe with increasing optical power might be explained by effects of the optical pump on the microwave resonator. 

\subsubsection{Detuning dependencies} 

\begin{figure*}[t]
\centering
\includegraphics[width=0.85\columnwidth]{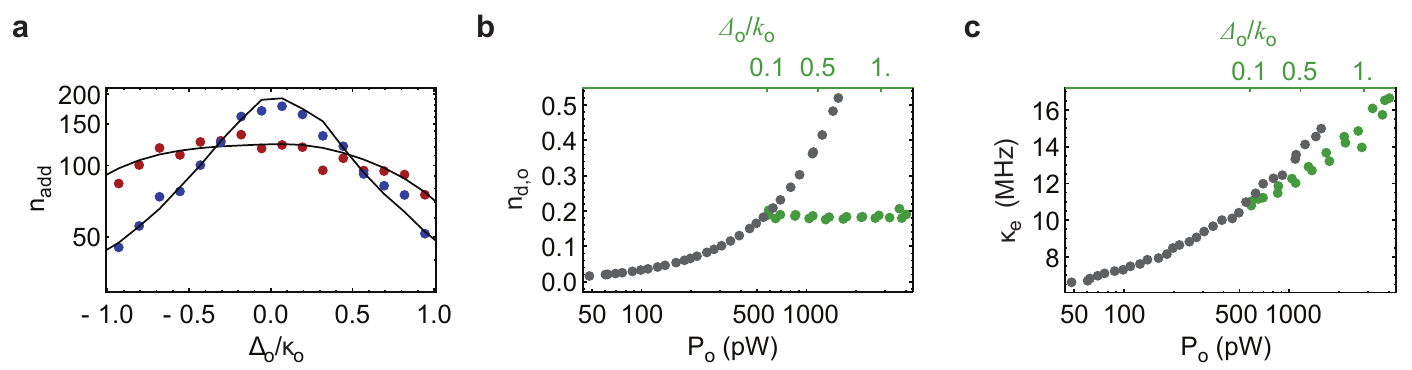}
\caption{\textbf{Transducer parameters vs detuning.} \textbf{a}, Electrical (red) and optical (blue) added noise at the transducer output with respect to the optical detuning ratio $\Delta_\text{o}/\kappa_\text{o}$. \textbf{b}, Drive photons in the optical cavity $n_\text{d,o}$ vs. optical pump power in the waveguide $P_\text{o}$ for constant detuning $\Delta_\text{o}/\kappa_\text{o} \approx 0.09$ (gray) and varying detuning with $n_\text{d,o} \approx 0.19$ (green). \textbf{c}, Microwave resonator linewidth $\kappa_{\text{e}}$ with respect to optical pump power. The resonator linewidth increase is stronger when the optical pump is applied near resonance (gray), compared to the detuning dependent case with lower and constant intracavity photon number $n_\text{d,o}$ (green). 
}\label{Fig_detuning}
\end{figure*}

The optomechanical gain~\cite{Massel2011a,Cohen2020} strongly depends on the pump detuning. Fig.~\ref{Fig_gain} in the main text shows the dependency of the total transduction $\zeta$ (Eq.~\ref{conversion0}), the conversion efficiency $\theta$ (Eq.~\ref{conversion01}), and the gain $\mathcal{G}$ (Eq.~\ref{conversion001}) of the transducer ($\omega \approx \omega_\text{m}'$) on the optical detuning for a constant optical intracavity photon number of $n_\text{d,o}=0.185 \pm 0.015$ and thereby a constant optomechanical coupling rate $G_\text{o}=g_\text{0,o} \sqrt{n_\text{d,o}}$. As described in the main text, the gain increases rapidly for vanishing detuning $\Delta_{\text{o}} \rightarrow 0$, while the conversion approaches 0 (Fig.~\ref{Fig_gain}). The corresponding numbers of the noise photons added to the electrical (red) and optical output (blue) are given in Fig.~\ref{Fig_detuning}. The electrical output noise is flattened compared to the optical output noise. The reason can be found in the phonon-photon coupling coefficient (\ref{coeff}d) because thermal mechanical occupation is the main noise source. The coefficient for coupling between phonons and the microwave (optical) photon output is proportional to the product of $G_{\text{e}} \chi_{\text{e}} $ ($G_{\text{o}} \chi_{\text{o}}$). While $G_{\text{e}} \chi_{\text{e}} $ is rather constant apart from changes in $\kappa_\text{e}$ due to optical heating and $G_{\text{o}}$ was kept constant vs. optical detunings by varying the pump power, $\chi_{\text{o}}$ and thereby $n_{\text{add,o}}$ depend strongly on the optical detuning and the optical output noise shows a peak around $\Delta_{\text{o}} = \omega_{\text{m}}$ ($\Delta_{\text{o}}/\kappa_{\text{o}} \approx $ 0.07). \\ 
In conjunction with the optical power sweep for constant optical detuning $\Delta_{\text{o}}/(2\pi) \approx 126\,\text{MHz}$ presented in the main text (Fig.~\ref{Fig_noise}c), it is then possible to compare heating rates for the same optical pump powers $P_{\text{o}}$ (optical powers in the waveguide) but different intracavity photon numbers $n_{\text{d,o}}$ (circulating power in the cavity) (Fig.~\ref{Fig_detuning}b). The deterioration of the microwave resonator linewidth $\kappa_{\text{e}}$ is a good indicator for optical heating. As can be seen in Fig.~\ref{Fig_detuning}c, $\kappa_{\mathrm{e}}$ rises faster with increasing optical pump power $P_{\text{o}}$ when the optical detuning is fixed and the intracavity photon number also increases (gray) compared to the case where the intracavity photon number is kept constant by varying $\Delta_{\text{o}}$ (green). At a pump power $P_{\text{o}} \approx 500\,\text{pW}$ and $\Delta_{\text{o}}/\kappa_{\text{o}} \approx  0.09$ we find a consistent $\kappa_{\text{e}}$ of 11.1\,MHz for both measurement sweeps. If $P_{\text{o}}$ is increased to 1500\,pW, $\kappa_{\text{e}}$ was broadened to 14.8\,MHz (33~\% increase) for constant detuning $\Delta_{\text{o}}/\kappa_{\text{o}} \approx 0.09$ ($n_{\text{d,o}} \approx 0.5$), but only to 12.8\,MHz (15~\% increase) for a different detuning $\Delta_{\text{o}}/\kappa_{\text{o}} \approx 0.65$ but a constant intracavity photon number $n_{\text{d,o}} \approx 0.19$. However, the fact that $\kappa_{\text{e}}$ increases also in the latter case of constant $n_{\text{d,o}}$ reveals that the intracavity photon number is not the only deciding factor for the heating rate. Similar trends can be found for the mechanical occupation. For $\Delta_{\text{o}}/\kappa_{\text{o}} \approx 0.09$ and $P_{\text{o}}=500\,\text{pW}$ ($n_{\text{d,o}} \approx 0.19$), we find again a consistent mechanical bath temperature $T_{\text{m}}$ of 0.65\,K for both measurement runs. If the pump power is increased to $P_o =1500\,\text{pW}$, $T_{\text{m}}$ rises by 23~\% to 0.85\,K ($n_{\text{d,o}} \approx 0.5$, $\Delta_{\text{o}}/\kappa_{\text{o}} \approx 0.09$), while it only heats up to 0.70\,K or 8~\%, if the intracavity photon number is kept at 0.19 ($\Delta_{\text{o}}/\kappa_{\text{o}} \approx 0.65$). We attribute the higher absorption heating by actual intracavity photons to the fact that the photonic crystal cavity has a direct physical connection to the microwave circuit by the mechanically compliant capacitors. If the optical photons do not enter the cavity but remain in the waveguide, there is only an indirect connection to the circuit via the silicon membrane and stray light. 

\end{document}